\documentclass[a4paper,twoside,12pt,onecolumn,aps,prd,tightenlines,secnumarabic,floatfix,nofootinbib,preprintnumbers,noshowpacs,notitlepage]{revtex4-1}
\pdfoutput=1

\usepackage[english]{babel}
\usepackage{amsmath,amssymb,bm,slashed}
\usepackage{graphicx}
\usepackage{booktabs}
\usepackage[sort&compress]{natbib}
\usepackage{xcolor}
\usepackage[normalem]{ulem}
\usepackage{hyperref}
\usepackage{cleveref}
\usepackage{subfigure} 
\definecolor{red}{rgb}{1.0, 0, 0}
\usepackage{verbatim}
\usepackage{setspace}
\usepackage{url}
\usepackage{slashed}
\usepackage{xspace}

\newcommand{\unit}[1]{\ensuremath{\mathrm{\,#1}}\xspace}

\newcommand{\tev}{\unit{TeV}}
\newcommand{\GeV}{\unit{GeV}}

\allowdisplaybreaks

\bibpunct{[}{]}{,}{n}{}{,}

\newcommand{\ev}[1]{\ensuremath{\left\langle #1 %
                     \right\rangle}} 

\makeatletter
\def\l@subsection#1#2{}

\makeatother

\begin{document}


\title{Symmetry Restored in Dibosons at the LHC?}

\author{Johann Brehmer$^{1}$}  \email[Email: ]{brehmer@thphys.uni-heidelberg.de}
\author{JoAnne Hewett$^{2}$}    \email[Email: ]{hewett@slac.stanford.edu}
\author{\\ Joachim Kopp$^{3}$}     \email[Email: ]{jkopp@uni-mainz.de}
\author{Thomas Rizzo$^{2}$}          \email[Email: ]{rizzo@slac.stanford.edu}
\author{Jamie Tattersall$^{4}$}   \email[Email: ]{tattersall@physik.rwth-aachen.de}
\affiliation{
  $^1$Institut f\"ur Theoretische Physik, Universit\"at Heidelberg, Germany \\
  $^2$SLAC National Accelerator Laboratory, Menlo Park, USA \\
  $^3$PRISMA Cluster of Excellence and Mainz Institute for Theoretical Physics,
    Johannes Gutenberg University, Mainz, Germany \\
  $^4$Institut f\"ur Theoretische Teilchenphysik und Kosmologie,
    RWTH Aachen, Germany }

\date{\today}
\preprint{MITP/15-046, SLAC-PUB-16319, TKK-15-15}


\begin{abstract}
  A number of LHC resonance search channels display an excess in the invariant mass region of 1.8\,--\,2.0 TeV. Among them is a $3.4\,\sigma$
  excess in the fully hadronic decay of a pair of Standard Model electroweak gauge bosons, in addition to 
  potential signals in the $HW$ and
  dijet final states. We perform a model-independent cross-section fit to the results of all ATLAS and CMS
  searches sensitive to these final states. We then interpret these results in the context of the Left-Right Symmetric Model, based on the extended gauge
  group $SU(2)_L\times SU(2)_R\times U(1)'$, and show that a heavy right-handed gauge boson $W_R$
  can naturally explain the current measurements with just a single coupling $g_R \sim 0.4$. In addition, we discuss a possible
  connection to dark matter.
\end{abstract}

\begin{flushright}

\end{flushright}

\maketitle

\tableofcontents

\section{Introduction}

  A number of searches for narrow resonances at the LHC now display peaks in the region of 1.8\,--\,2.0 TeV. Most prominently, a recent 
  ATLAS search \cite{Aad:2015owa} for a resonance that decays to a pair of standard model (SM) gauge bosons contains a
  local excess of 3.4\,$\sigma$ (2.5\,$\sigma$ global) in the $WZ$ final state at approximately 2~TeV. Since the search is fully
  hadronic, there is only a limited ability to accurately distinguish gauge bosons, and consequently many of the events can also
  be interpreted as a $ZZ$ or $WW$ resonance, leading to excesses of 2.9\,$\sigma$ and 2.6\,$\sigma$ in these channels respectively.
  
  Interest is further piqued when one analyses other resonance searches in a similar mass range. At approximately
  1.8~TeV, both CMS \cite{Khachatryan:2015sja} and ATLAS \cite{Aad:2014aqa} observe excesses in the dijet distributions
  with a significance of 2.2\,$\sigma$ and 1\,$\sigma$, respectively. In addition, a CMS search for resonant $HW$ production
  \cite{CMS:2015gla} shows a 2.1\,$\sigma$ excess,
  and another CMS search \cite{Khachatryan:2014gha} for a pair of vector bosons, but this time with a leptonically tagged $Z$,
  finds a 1.5\,$\sigma$ excess, both at approximately 1.8~TeV. With the exception of the three ATLAS selections in gauge boson pair production,
  all these possible signals are completely independent.

  We analyze these excesses and perform a general cross-section fit to the data. It is of course important 
  to not only look at possible signals in the distributions, and for that reason we include all relevant searches that may be sensitive to
  the same final states as those showing anomalies. These include diboson analyses from both ATLAS \cite{Aad:2015ufa,Aad:2014xka,Aad:2015yza} 
  and CMS \cite{Khachatryan:2014hpa,Khachatryan:2015ywa,Khachatryan:2015bma}. In addition, many models that explain an 
  excess of events in the dijet distribution with a charged mediator will also lead to a peak in a $tb$ resonance search.
  Consequently, we also include the most sensitive searches for this particular final state from both 
  ATLAS \cite{Aad:2014xra,Aad:2014xea} and CMS \cite{Chatrchyan:2014koa} in our study.
  
  Combining all searches, we provide best fit signal cross sections for each of the final states analyzed in 
  order to guide the model building process. In particular we find that the vector boson pair production searches
  are best described by a $WZ$ or $ZZ$ final state, while $WW$ is disfavoured via limits from semi-leptonic searches. In 
  the associated Higgs production fit, we find preference for the $HW$ final state, since an excess here is only seen in
  the single-lepton analysis. Finally, there is good agreement for a dijet signal between ATLAS and CMS, but nothing has 
  been observed so far in the $tb$ final state.
  
  In order to explain these excesses we focus on the so-called Left-Right Symmetric Model
  (LRM)~\cite{Pati:1974yy,Mohapatra:1974hk,Mohapatra:1974gc,Senjanovic:1975rk,Mohapatra:1986uf}.
  The LRM is based on the low-energy gauge group 
  $SU(2)_L\times SU(2)_R\times U(1)'$ that can arise for example from an $SO(10)$ or $E6$ GUT~\cite{Langacker:1980js,Hewett:1988xc}.
  Since there is a new
  $SU(2)$ gauge group in addition to the SM, the spectrum now contains both a $Z_R$ and charged $W^{\pm}_R$ bosons.
  In general the model predicts $m_{Z_R}>m_{W^{\pm}_R}$, which is why we explain the various signals via resonant $W^{\pm}_R$
  production and decay. By extending our cross-section fit to the LRM parameters, we find a region that can explain
  the excesses while avoiding all current exclusion bounds. We then examine where our fit suggests the mass of the
  $Z_R$ should be and the discovery potential of the LHC for this state.
  
  In a next step we also analyze the question of how such a model may be able to simultaneously explain dark matter (DM). 
  If the $W_R$ is to mediate DM annihilation in the early universe the simplest scenario we can find is 
  to include a new charged and neutral particle that would be relatively close in mass in order to co-annihilate effectively. 
  Alternatively, we offer the concrete example that DM annihilation is mediated through the $Z_R$ and the dark matter candidate
  is the neutral component of a new fermionic doublet. In this case, we see that resonant annihilation is required for the correct
  relic density and thus we predict the DM mass is $\sim \frac{1}{2}m_{Z_R}$.

  This paper begins in Sec.~\ref{sec:measurements} with a review of the different experimental studies under consideration and
  generic cross-section fits. In Sec.~\ref{sec:LRM} we interpret the results in the LRM, followed by a discussion of a potential
  link to dark matter in Sec.~\ref{sec:DM}. Finally, we give our conclusions in Sec.~\ref{sec:conclusions}.

\section{Experimental measurements and cross-section fits}
\label{sec:measurements}

In this section we give an overview over the ATLAS and CMS searches in which resonances in the mass range 1.8\,--\,2.0~TeV have
been observed, or which provide constraints in the same or closely related final states. It is important
to note that due to resolution, jet energy scale uncertainties and the limited number of events currently present in the excesses, peaks
at 1.8\,--\,2.0~TeV can easily be compatible. Even though the most significant excess is observed in
the ATLAS diboson search~\cite{Aad:2015owa} at an invariant mass of 2.0~TeV, the majority of excesses
are reported closer to an invariant mass of approximately 1.8~TeV. This is why we assume a resonance in the vicinity
of 1.8\,--\,1.9~TeV in this analysis.

We analyze the results of these searches by performing cross-section fits in each channel individually.
For this, we sum bins around this mass region according to the experimental width and jet energy scale
of the particular study. We then perform a cut-and-count analysis on the event numbers in these enlarged signal regions.
As input information we take into account the number of observed events, expected backgrounds, efficiencies,
and systematic uncertainties, as published by ATLAS and CMS. Where necessary information
is missing from the publications, we use estimations to the best of our knowledge. In App.~\ref{sec:appendix} we
list the data we use in more detail.

For the actual fit, we follow a frequentist approach. In the absence of systematic uncertainties, the number of
events observed in the signal region, $n$,
follows a Poisson distribution
\begin{equation}
  p (n) = \frac {\mu^n \exp(-\mu)} {n!} \,.
\end{equation}
Here $\mu = b + \sigma_s \cdot BR \cdot L \cdot \varepsilon$ is the number of expected events, including the number of expected
background events $b$, and the product of the signal production cross section $\sigma_s$, branching
ratios $BR$, integrated luminosity $L$, and efficiency as well as acceptance factors $\varepsilon$.
The systematic uncertainties on the background prediction $b$ and on the signal efficiency $\varepsilon$ are approximated by
Gaussian distributions. These nuisance parameters are marginalized. We neglect correlations between the different systematic
uncertainties.

We then calculate a $p$ value for each signal cross section
in each different final state, using at least $10\,000$ pseudo-experiments. This gives us the significance of the deviation
from the SM, best fit points, and confidence regions
at $68 \,\%$ and $90 \,\%$ CL. We also calculate upper limits in a modified frequentist approach (CLs method \cite{0954-3899-28-10-313}).
In some of the signal regions, our simple cut-and-count strategy leads to model limits that are stronger than those presented by the experiments.
In this case, the systematical background error and the signal efficiencies
were rescaled to find agreement. Since many of our input parameters involve rough estimations, we have checked
that the final results are stable under the variation of these input values.

In a next step, we perform combined fits to all studies that contribute to a particular final state. For the combination of
individual $p$ values we use Fisher's method. Again, we present the results in terms of $p_0$ values, best fit points,
and confidence regions at $68 \,\%$ and $90 \,\%$ CL.

\subsection{Vector boson pair production}

\begin{table}[ht]
\renewcommand{\arraystretch}{1.}
\begin{tabular*}{1.0\textwidth}{@{\extracolsep{\fill} }lcccc} 

  \multicolumn{3}{l}{\textbf{$WW$ resonance analyses}}   \\ \hline
    Analysis     	        		& Expected       &   Observed       &  Excess	     		& Fitted cross	   \\
						& 95\% CLs [fb]  &   95\% CLs [fb]  &  significance [$\sigma$]  & section [fb] 	   \\ \hline\hline
 ATLAS hadronic \cite{Aad:2015owa}		&     11.1	 &  	19.6	    &    1.2		 	&  	5.2	     \\
 CMS hadronic \cite{Khachatryan:2014hpa}    	&     12.2	 &  	17.9	    &     1.0		 	&  	6.0	     \\
 ATLAS single lepton \cite{Aad:2015ufa}		&     6.4	 &  	5.9	    &     0.0		 	&  	0.0	     \\
 CMS single lepton \cite{Khachatryan:2014gha}	&     7.2	 &  	8.1	    &    0.3 		 	&  	1.2	     \\
 \hline 
\end{tabular*}
\caption{Expected and observed limits, significance of any excess present, and the potential signal 
cross section for $WW$ pair production at 1.8~TeV. Values may differ from those reported in the experimental papers since we integrate 
over signal regions in the vicinity of 1.8~TeV depending on the resolution and binning of the selected 
analysis.}
\label{tab:ww_analyses}
\end{table}

\begin{table}[ht]
\renewcommand{\arraystretch}{1.}
\begin{tabular*}{1.0\textwidth}{@{\extracolsep{\fill} }lcccc} 

  \multicolumn{3}{l}{\textbf{$WZ$ resonance analyses}}   \\ \hline
    Analysis     	        		& Expected       &   Observed       &  Excess	     		& Fitted cross	   \\
						& 95\% CLs [fb]  &   95\% CLs [fb]  &  significance [$\sigma$]  & section [fb] 	   \\ \hline\hline
ATLAS hadronic \cite{Aad:2015owa}		&     14.2	 &  	25.8	    &     	1.3	 	&  	6.9	     \\
CMS hadronic \cite{Khachatryan:2014hpa}		&     11.9	 &  	17.5	    &     1.0		 	&  	5.8	     \\
ATLAS single lepton \cite{Aad:2015ufa}		&     13.2	 &  	12.4	    &     0.0		 	&  	0.0	     \\
CMS single lepton \cite{Khachatryan:2014gha}	&     14.9	 &  	16.8    &     0.3		 	&  	2.4	     \\
ATLAS double lepton \cite{Aad:2014xka}		&     13.8	 &  	20.5	    &     0.3		 	&  	2.9	     \\
CMS double lepton \cite{Khachatryan:2014gha}	&     14.4	 &  	27.4	    &     1.5		 	&  	10.0	     \\
 \hline 
\end{tabular*}
\caption{Expected and observed limits, significance of any excess present, and the potential signal 
cross section for $WZ$ production at 1.8~TeV. Values may differ from those reported in the experimental papers since we integrate 
over signal regions in the vicinity of 1.8~TeV depending on the resolution and binning of the selected 
analysis.}
\label{tab:wz_analyses}
\end{table}

\begin{table}[ht]
\renewcommand{\arraystretch}{1.}
\begin{tabular*}{1.0\textwidth}{@{\extracolsep{\fill} }lccccc} 

  \multicolumn{3}{l}{\textbf{$ZZ$ resonance analyses}}   \\ \hline
    Analysis     	        		& Expected       &   Observed       &  Excess	     		& Fitted cross	   \\
						& 95\% CLs [fb]  &   95\% CLs [fb]  &  significance [$\sigma$]  & section [fb] 	   \\ \hline\hline
 ATLAS hadronic \cite{Aad:2015owa}		&    9.7	 &  	25.2	    &     2.4		 	&  	8.1	     \\
 CMS hadronic \cite{Khachatryan:2014hpa}	&    11.7	 &  	17.1	    &     1.0		 	&  	5.7	     \\
 ATLAS double lepton \cite{Aad:2014xka}		&    6.7	 &  	10.0	    &     0.3		 	&  	1.4	     \\
 CMS double lepton \cite{Khachatryan:2014gha}	&    7.0	 &  	13.4	    &     1.5		 	&  	4.9	     \\
 \hline 
\end{tabular*}
\caption{Expected and observed limits, significance of any excess present, and the potential signal 
cross section for $Z$ pair production at 1.8~TeV. Values may differ from those reported in the experimental papers since we integrate 
over signal regions in the vicinity of 1.8~TeV depending on the resolution and binning of the selected 
analysis.}
\label{tab:zz_analyses}
\end{table}

\begin{figure}
 \centering
 \includegraphics[width=0.49\textwidth]{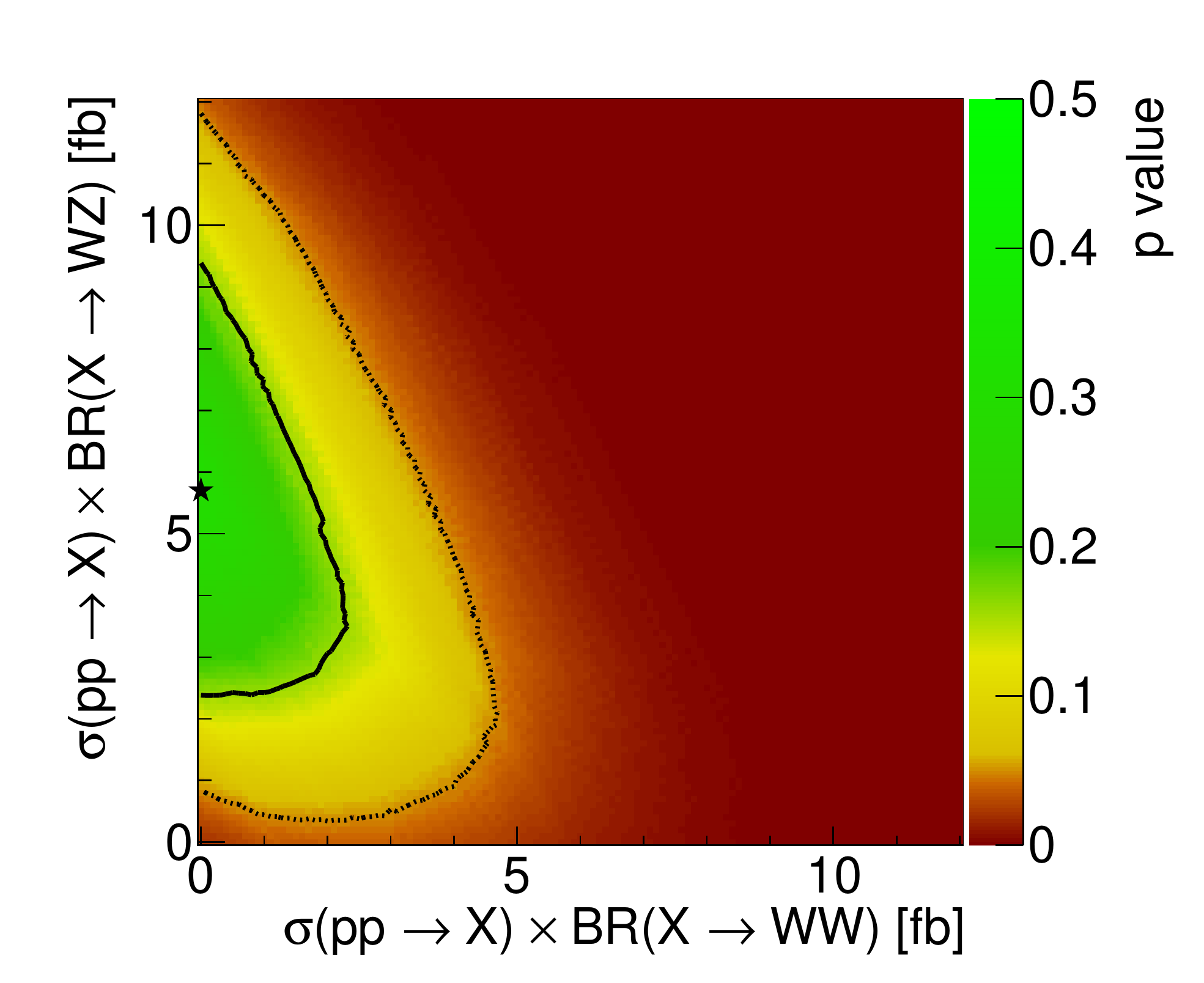}
 \includegraphics[width=0.49\textwidth]{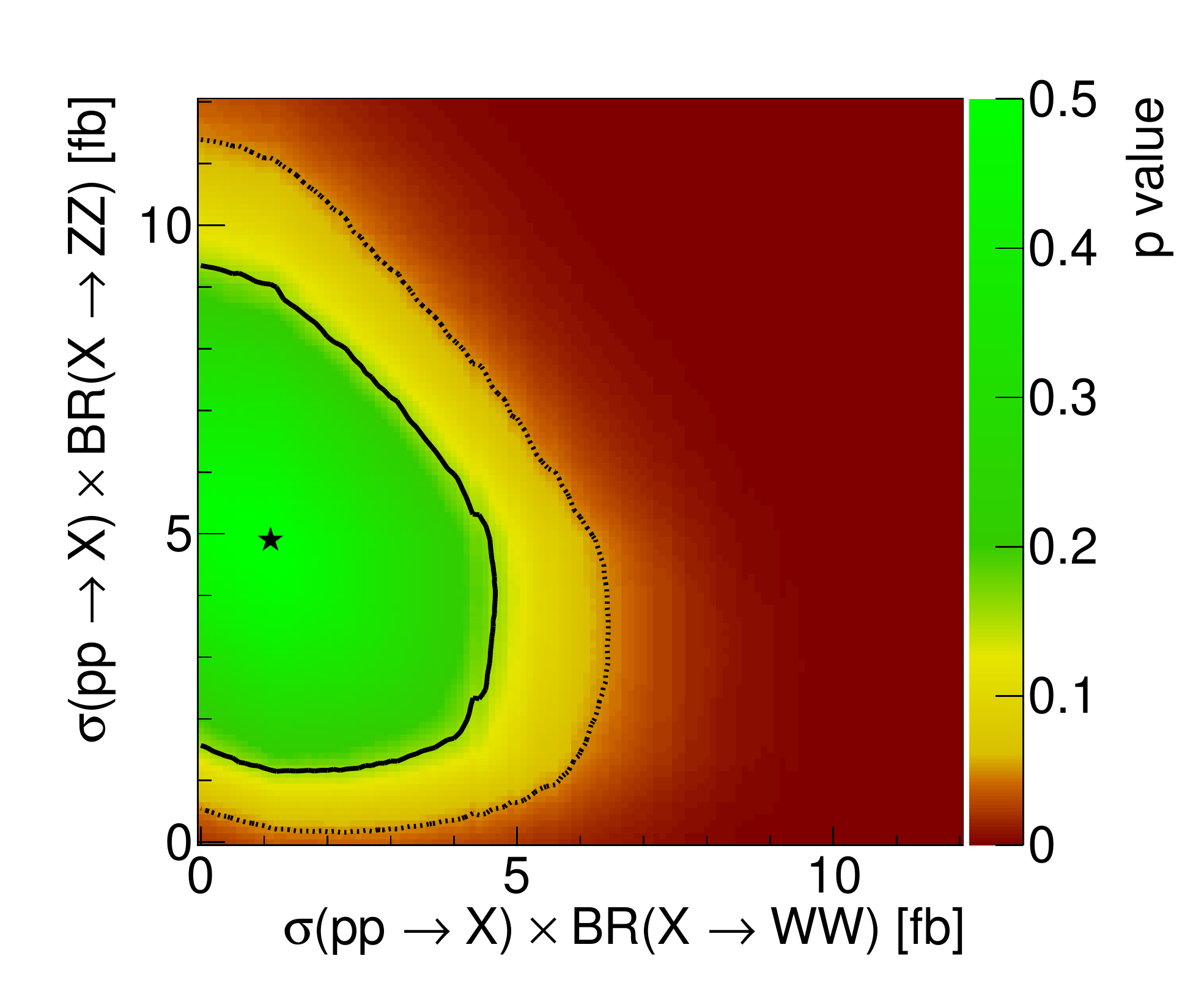}
 \caption{$p$ values for a 1.8~TeV resonance decaying into a combination of $WZ$ and $WW$ pairs (left), or 
 $ZZ$ and $WW$ pairs (right). The 68\% (90\%) preferred region is shown with a solid (dotted) line. The star represents the best fit point.}
 \label{fig:VecVec}
\end{figure}

The relevant searches targeting the resonant production of a pair of vector bosons either require both bosons to decay
hadronically~\cite{Aad:2015owa,Khachatryan:2014gha} or one boson leptonically~\cite{Aad:2015ufa,Aad:2014xka,Khachatryan:2014gha}. 
Unfortunately, branching ratio suppression dictates that searches where both bosons decay leptonically are currently uncompetitive.

In this set of analyses, the ATLAS fully hadronic search presently contains the largest single excess with $3.4\,\sigma$ 
reported by the experiment for an invariant mass of 2.0~TeV. Our cut-and-count analysis focuses on a slightly lower invariant mass window,
leading to a reduced peak significance of $2.4\,\sigma$, see Tab.~\ref{tab:zz_analyses}. In order 
to fit the different final states we make use of the fact that the analysis has a mild discrimination 
between $W$ and $Z$ bosons based on the invariant
mass of a reconstructed `fat jet'. This leads to a slight preference for either a $WZ$ or $ZZ$ final state to explain the excess
present with a cross section of $\sim 7$\,--\,8~fb, see Tab.~\ref{tab:wz_analyses} and \ref{tab:zz_analyses}. Since a smaller peak is 
seen in the purely $WW$ channel, a smaller 
cross section of $\sim 5$~fb is found for this channel, see Tab.~\ref{tab:ww_analyses}. We again note that many of the events seen in the excess
region are shared between all three final states; thus the total cross section in the excess region is substantially
smaller than if we simply sum the three fitted cross sections together.

The CMS fully hadronic analysis search is very similar and also finds an excess in the same mass range, although with a slightly
smaller signal of $\sim 1\,\sigma$, see Tab.~\ref{tab:ww_analyses}\,--\,\ref{tab:zz_analyses}. In this analysis no real
preference is seen for any of the different final states and a cross section of $\sim 6$~fb fits all three equally well.

More discrimination of the final states is available by using the semi-leptonic searches. Here, the one-lepton analyses
require that a $W$ boson is present while the two-lepton searches reconstruct at least one $Z$ boson. If we first examine the
one-lepton searches we see that no excess is seen by ATLAS \cite{Aad:2015ufa}, while CMS \cite{Khachatryan:2014gha} 
only has a very mild excess of $\sim 0.3$\,$\sigma$. These searches place a significant constraint on the $WW$ final state 
with ATLAS and CMS giving limits around 6\,--\,8~fb at 95\% CLs. Removing one $W$ for the $WZ$ final state relaxes this
limit to 12~fb, see Tab.~\ref{tab:wz_analyses}. 

In the two-lepton case, CMS \cite{Khachatryan:2014gha} sees an excess of 1.5\,$\sigma$ which leads to a
fitted cross section of 10~fb if we assume a pure $WZ$ final state, see Tab.~\ref{tab:wz_analyses}, or 5~fb for a 
pure $ZZ$ final state, see Tab.~\ref{tab:zz_analyses}. The ATLAS search has similar sensitivity \cite{Aad:2014xka} but only observes
a very small excess of $0.3\,\sigma$. 

\begin{figure}
 \centering
 \includegraphics[width=0.49\textwidth]{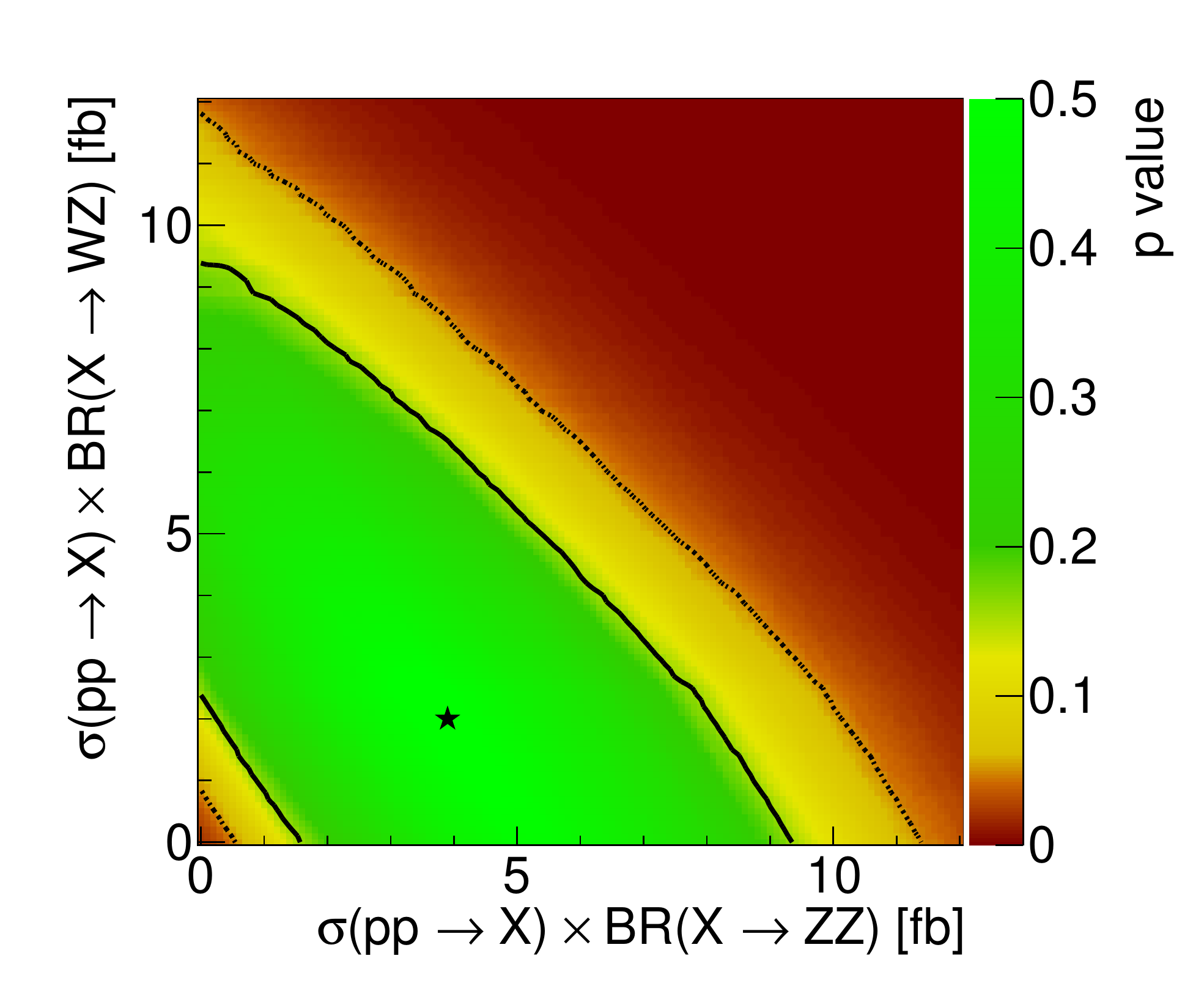}
 \includegraphics[width=0.49\textwidth]{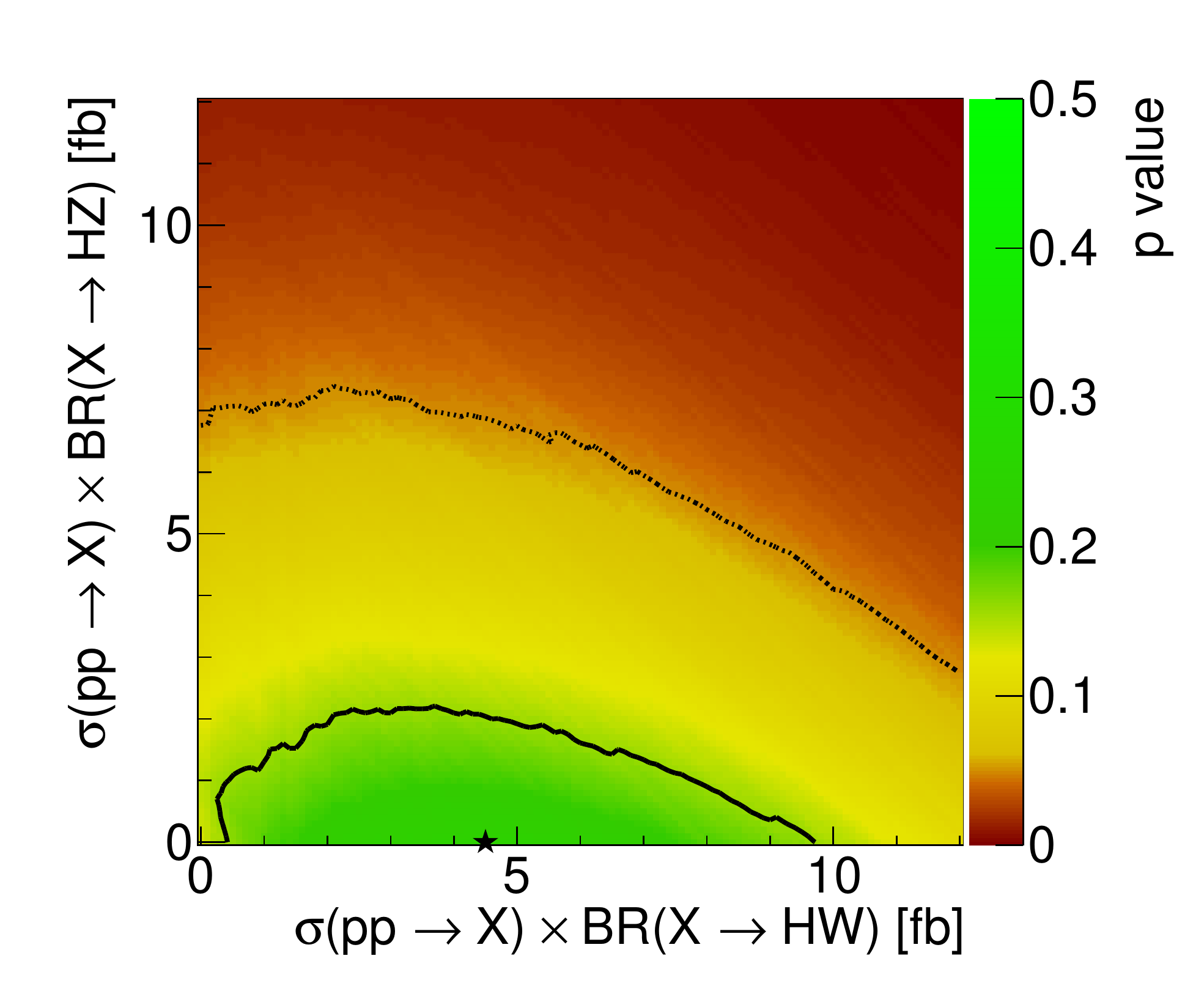}
 \caption{$p$ values for a 1.8~TeV resonance decaying into a combination of $WZ$ and $ZZ$ (left), or
 $HZ$ and $HW$ pairs (right). The 68\% (90\%) preferred region is shown with a solid (dotted) line. The star represents the best fit point.}
 \label{fig:HiggsVec}
\end{figure}

In Fig.~\ref{fig:VecVec} and Fig.~\ref{fig:HiggsVec} (left panel) we show the combined cross-section fit to all 
channels. Fig.~\ref{fig:VecVec} shows that both $WZ$ and $ZZ$ final states fit the data well with a cross section of
$\sim 5$~fb. However, we also see that a pure $WW$ signal is disfavoured and could only describe the
data in combination with another signal. The reason that the $WW$ explanation is disfavoured is two-fold. First, 
the ATLAS and CMS single lepton analyses set an upper limit around $6$~fb at 95\% CLs, but a cross section 
of this magnitude is required to fit the hadronic excesses. In addition, the CMS dilepton search~\cite{Khachatryan:2014gha}
has a small excess that this channel cannot explain.

Finally the left panel of Fig.~\ref{fig:HiggsVec} shows that an equally good fit is possible with contributions from both a $WZ$ and $ZZ$
final state, i.\,e.\ requiring both a charged and neutral resonance with similar mass. In combination, the standard model is
disfavoured by $\sim 2\,\sigma$ when all vector boson pair production
channels are included.

\subsection{Associated vector-Higgs production}
\label{sec:vec_higgs}

 \begin{table}[ht]
\renewcommand{\arraystretch}{1.}
\begin{tabular*}{1.0\textwidth}{@{\extracolsep{\fill} }lccccc} 

  \multicolumn{3}{l}{\textbf{$HW$ resonance analyses}}   \\ \hline
    Analysis     	        					& Expected       &   Observed       &  Excess	    		 & Fitted cross	   \\
									& 95\% CLs [fb]  &   95\% CLs [fb]  &  significance [$\sigma$]   & section [fb] 	   \\ \hline\hline
 ATLAS $b\overline{b}+(\ell\ell,\nu\ell,\nu\nu)$ \cite{Aad:2015yza}	&     33.0	 &  	30.0	    &     0.0		 	&  	0.0	     \\
 CMS $b\overline{b}+\nu\ell$ \cite{CMS:2015gla}    			&     18.6	 &  	44.4	    &     1.9		 	&  	15.8     \\
 CMS $\tau^+\tau^-$ + hadronic vector \cite{Khachatryan:2015ywa}    	&     36.1	 &  	36.1	    &     0.0		 	&  	0.0	     \\
 CMS hadronic Higgs  \cite{Khachatryan:2015bma}				&     12.5	 &  	13.2	    &     0.1		 	&  	1.0	     \\
 \hline 
\end{tabular*}
\caption{Expected and observed limits, significance of any excess present, and the potential signal 
cross section for associated production of a Higgs and a $W$ at 1.8~TeV. Values may differ from those reported in the experimental papers since we integrate 
over signal regions in the vicinity of 1.8~TeV depending on the resolution and binning of the selected 
analysis.}
\label{tab:hw_analyses}
\end{table}
 
 \begin{table}[ht]
\renewcommand{\arraystretch}{1.}
\begin{tabular*}{1.0\textwidth}{@{\extracolsep{\fill} }lccccc} 

  \multicolumn{3}{l}{\textbf{$HZ$ resonance analyses}}   \\ \hline
    Analysis     	        					& Expected       &   Observed       &  Excess	    		 & Fitted cross	   \\
									& 95\% CLs [fb]  &   95\% CLs [fb]  &  Significance [$\sigma$]   & section [fb] 	   \\ \hline\hline
 ATLAS $b\overline{b}+(\ell\ell,\nu\ell,\nu\nu)$ \cite{Aad:2015yza}	&     15.0	 &  	14.0	    &     0.0		 	&  	0.0	     \\
 CMS $\tau^+\tau^-$ + hadronic vector \cite{Khachatryan:2015ywa}	&     	31.8	 &  	31.8	    &     0.0		 	&  	0.1	     \\
 CMS hadronic Higgs  \cite{Khachatryan:2015bma}				&     	12.2	 &  	12.9	    &     0.1		 	&  	1.0	     \\
 \hline 
\end{tabular*}
\caption{Expected and observed limits, significance of any excess present, and the potential signal 
cross section for associated production of a Higgs and a vector boson at 1.8~TeV. Values may differ from those reported in the experimental papers since we integrate 
over signal regions in the vicinity of 1.8~TeV depending on the resolution and binning of the selected 
analysis.}
\label{tab:hz_analyses}
\end{table} 

The experimental searches for a resonance that decays into a $HV$ final state are varied. ATLAS looks for a Higgs
that produces a $b\bar{b}$ pair with the $W$ or $Z$ probed leptonically $(\ell\ell$, $\nu\ell$ or $\nu\nu)$~\cite{Aad:2015yza}.
CMS has a similar search for $H\to b\bar{b}$ but only examines the leptonic $W$ channel~\cite{CMS:2015gla}. To probe
vector bosons more generally, CMS has a fully hadronic search~\cite{Khachatryan:2015bma}, but this has a limited ability
to discriminate between $W$ and $Z$. Finally there is a CMS search for $H\to \tau^+\tau^-$, again with a hadronic reconstruction
of the vector boson~\cite{Khachatryan:2015ywa}.

Out of these searches, only the CMS study with a leptonic $W$ displays an significant excess with $\sim2\,\sigma$ at 
1.8\,--\,1.9~TeV. Interpreting this as a resonance that decays to $HW$ leads to a fitted cross section of 
16~fb, see Tab.~\ref{tab:hw_analyses}. However, the fully hadronic CMS search is slightly in tension with
this result as it reports a limit of 13~fb on the same final state.
In the analyses that are sensitive to $HZ$ production, no significant excesses are 
seen, see Tab.~\ref{tab:hz_analyses}. For this final state, the ATLAS semi-leptonic \cite{Aad:2015yza} and 
CMS fully hadronic~\cite{Khachatryan:2015bma} searches have similar sensitivities and set a 95\% CLs limit of
14~fb and 13~fb respectively.

Combining all these searches into a single fit, we plot the preferred cross sections for $HW$ and $HZ$ productions 
in the right panel of Fig.~\ref{fig:HiggsVec}. We find that the best fit point has a cross section for $HW$ production of $\sim 5$~fb,
but is also compatible with the SM background at $\sim 1\,\sigma$. Since the only excess is seen in 
a channel compatible with a $W$ in the final state, there is no evidence in the data for a signal in the $HZ$ channel.

\subsection{Dijet production}

Both the dijet search by ATLAS \cite{Aad:2014aqa} and that by CMS \cite{Khachatryan:2015sja} see an excess in 
the invariant mass distribution around 1.8~TeV. In our analysis, we find the excess in CMS is slightly
more significant at 1.9\,$\sigma$ compared to 1.5\,$\sigma$ in ATLAS. However, since the CMS analysis
is slightly more sensitive to a signal, the fitted cross section is actually smaller at $\sim 90$~fb, compared 
to ATLAS with $\sim$100~fb, see Tab.~\ref{tab:jj_analyses}. In any case, the two signals seen by both experiments 
are remarkably similar and the combined best-fit signal cross section of $\sim 90$~fb can be seen in Fig.~\ref{tab:tb_analyses}.
 
 \begin{table}[ht]
\renewcommand{\arraystretch}{1.}
\begin{tabular*}{1.0\textwidth}{@{\extracolsep{\fill} }lccccc} 
  \multicolumn{3}{l}{\textbf{Dijet resonance analyses}}   \\ \hline
    Analysis     	        		& Expected       &   Observed       &  Excess	    		 & Fitted cross	   \\
						& 95\% CLs [fb]  &   95\% CLs [fb]  &  Significance [$\sigma$]   & section [fb] 	   \\ \hline\hline
 ATLAS dijet \cite{Aad:2014aqa}		&    131	 &  	217	    &     1.5		 	&  	101	     \\
 CMS dijet \cite{Khachatryan:2015sja}    	&     92	 &  	173	    &     1.9		 	&  	90	     \\
 \hline 
\end{tabular*}
\caption{Expected and observed limits, significance of any excess present, and the potential signal 
cross section for dijet production at 1.8~TeV. Values may differ from those reported in the experimental papers since we integrate 
over signal regions in the vicinity of 1.8~TeV depending on the resolution and binning of the selected 
analysis.}
\label{tab:jj_analyses}
\end{table}

\subsection{Associated top-bottom production}
 \label{sec:tb_searches}
 
  \begin{table}[ht]
\renewcommand{\arraystretch}{1.}
\begin{tabular*}{1.0\textwidth}{@{\extracolsep{\fill} }lccccc} 

  \multicolumn{3}{l}{\textbf{$t-b$ resonance analyses}}   \\ \hline
    Analysis     	        		& Expected       &   Observed       &  Excess	    		 & Fitted cross	   \\
						& 95\% CLs [fb]  &   95\% CLs [fb]  &  Significance [$\sigma$]   & section [fb] 	   \\ \hline\hline
 ATLAS hadronic $t$ \cite{Aad:2014xra}		&   155		 &  	203	    &     0.6		 	&  	31	     \\
 ATLAS leptonic $t$ \cite{Aad:2014xea}		&    138	 &  	101	    &     0.0		 	&  	0	     \\
 CMS leptonic $t$ \cite{Chatrchyan:2014koa}    	&    76		 & 	67	    &     0.0		 	&  	0	     \\
 \hline 
\end{tabular*}
\caption{Expected and observed limits, significance of any excess present, and the potential signal 
cross section for resonant production of a top and bottom quark at 1.8~TeV. Values may differ from those reported in the experimental papers since we integrate 
over signal regions in the vicinity of 1.8~TeV depending on the resolution and binning of the selected 
analysis.}
\label{tab:tb_analyses}
\end{table}

The final analyses that we consider are the studies that look for the resonant production of a $tb$
final state, see Tab.~\ref{tab:tb_analyses}. ATLAS has two searches that focus on this signature, one that looks for the hadronic 
decay of the $t$ \cite{Aad:2014xra} and another that considers the leptonic channel \cite{Aad:2014xea}. 
On the CMS side, only one study exists, concentrating on the leptonic decay \cite{Chatrchyan:2014koa}.
At the current time, only the ATLAS hadronic search contains a small excess in the region of 1.8~TeV. 
However, since this search was expected to have the poorest sensitivity, it is likely that this is 
purely a statistical fluctuation.

The strongest bound comes from the CMS leptonic search with an upper limit on the cross section
of 70~fb at 95\% CLs. However, we should also pay particular attention to the ATLAS leptonic search
since this heavily influences our final model fits and interpretations. The reason is that at 1.8~TeV, the 
search records a signal 1.8~$\sigma$ smaller than expected. We hesitate to call this an `underfluctuation'
because the search systematically measures a cross section 2~$\sigma$ less than the background prediction
across the whole mass probed (500\,--\,3000~GeV). Due to the construction of the CLs limit setting procedure,
which weights the likelihood according to the agreement of the signal with the background,
the systematically high background prediction does not result in a hugely significant shift in the 95\% CLs limit
(expected limit: 138~fb, observed limit: 101~fb).

\begin{figure}
 \centering \vspace{-0.0cm}
 \includegraphics[width=0.49\textwidth]{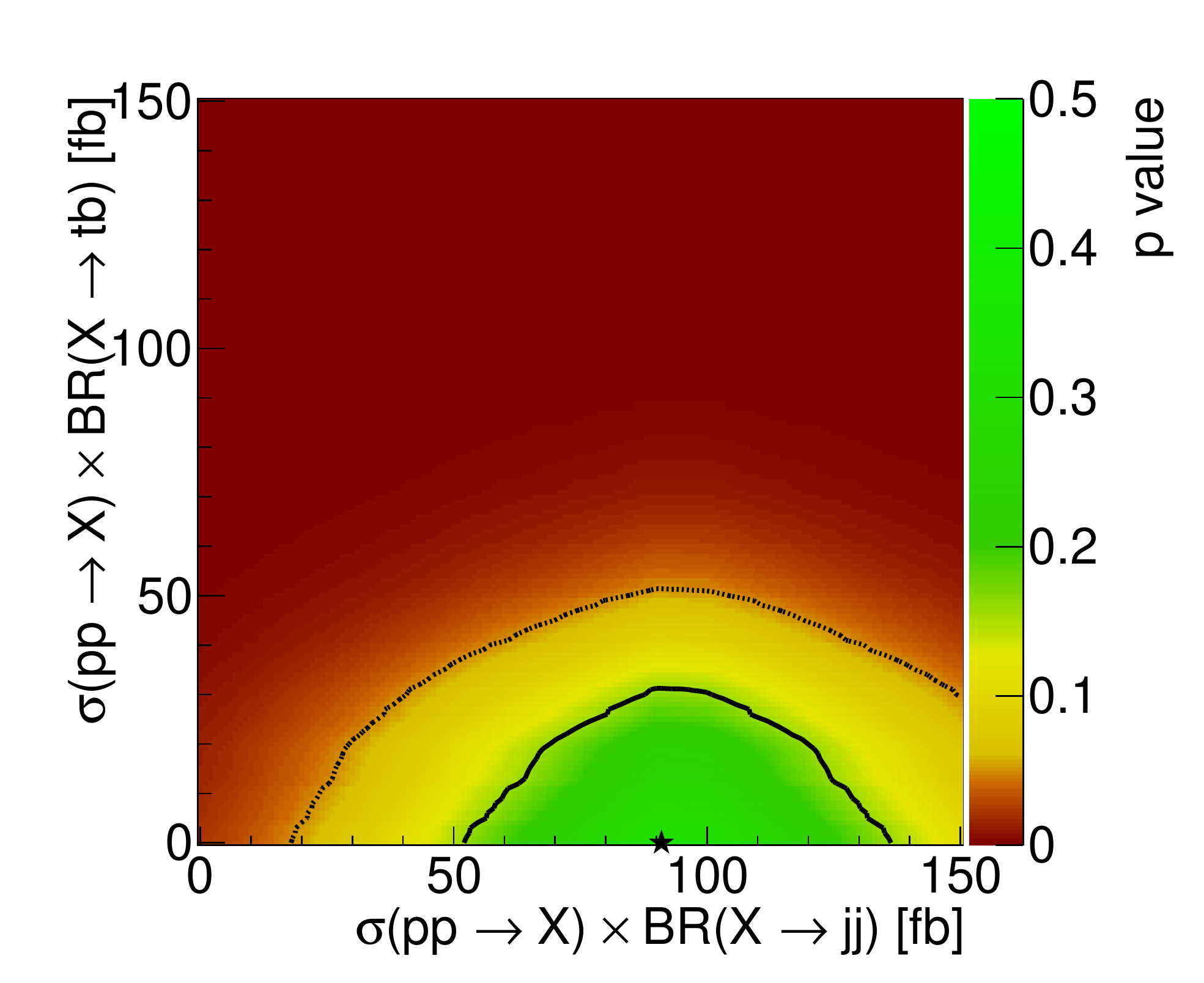}
 \includegraphics[width=0.49\textwidth]{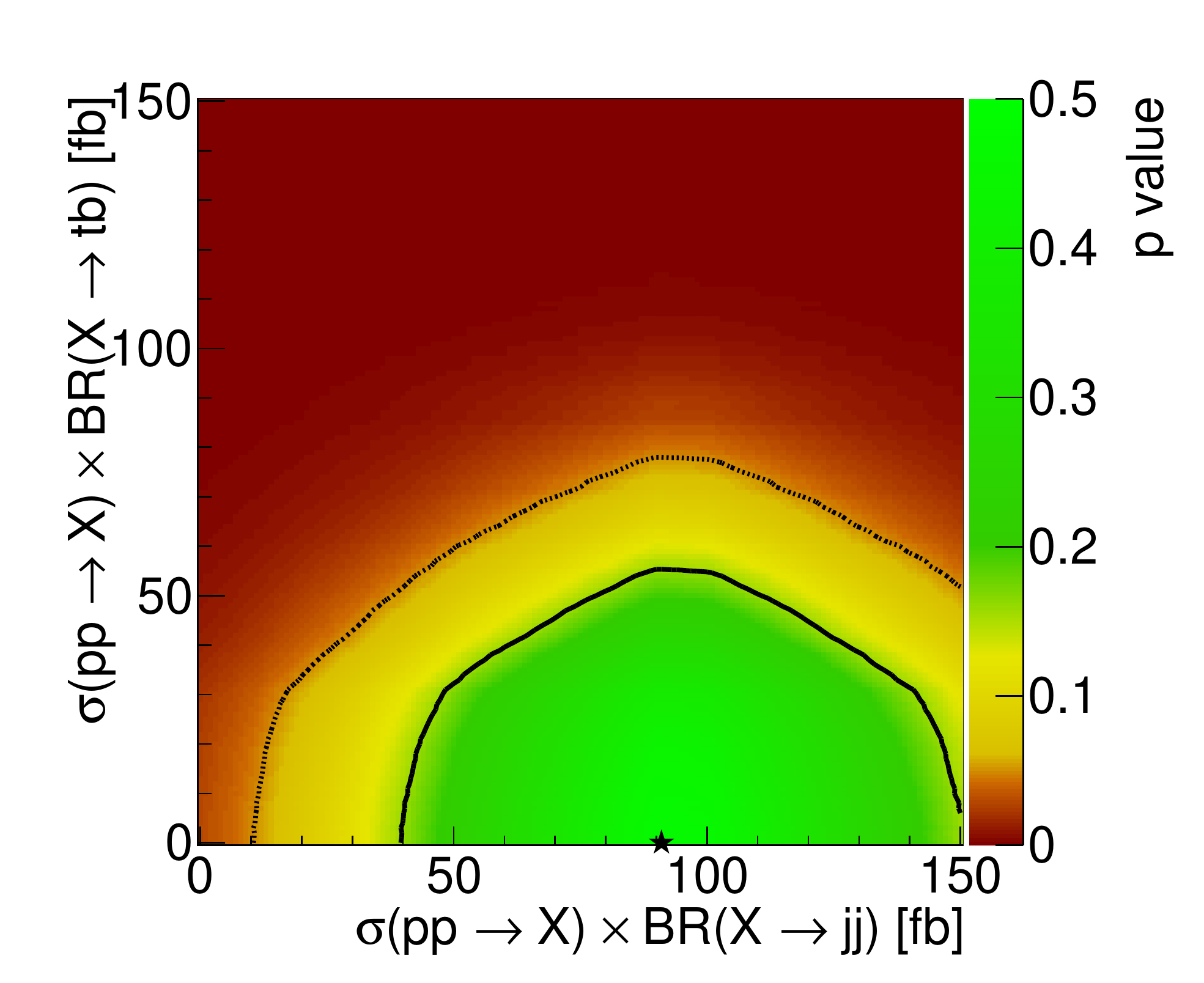}
 \caption{$p$ values for a 1.8 TeV resonance decaying into a combination of $jj$ and $tb$. The 
 68\% (90\%) preferred region is shown with a solid (dotted) line.  The star represents the best fit point.  Left: including the ATLAS
 leptonic $tb$ search that displays a systematically $\sim$1.8\,$\sigma$ improved exclusion. Right: 
 excluding the ATLAS leptonic $tb$ search from the fit.
 \label{fig:dijet_tb}}
\end{figure}

However, in our cross-section fit, the search has a far greater effect on the overall $p$ value, since
any additional signal predicted in this channel will be heavily penalized. For this reason, we perform 
two fits to the $tb$ final state, one which includes the ATLAS leptonic search, shown in the left panel of Fig.~\ref{fig:dijet_tb},
and one without, see the right panel of Fig.~\ref{fig:dijet_tb}. In both plots we see that the best fit point is found without 
a signal present. However, including the leptonic ATLAS search results in a 1\,$\sigma$ allowed cross section of
$\sim 30$~fb but removing the search allows this to increase to $\sim 60$~fb in combination with the dijet result.

\subsection{Channel comparison}

\begin{table}[ht]
\renewcommand{\arraystretch}{1.}
\begin{tabular*}{.8\textwidth}{@{\extracolsep{\fill} }lcc} 
  \multicolumn{3}{l}{\textbf{Fitted cross sections}}   \\ \hline
  Process      	        						& Fitted cross 	 & Upper bound    	   \\ 
        	        				           		& section [fb]	 & (90\% CL)    	   \\ \hline\hline
 $pp \to X \to WZ^1$    						& $5.7^{+3.6}_{-3.3}$   & 11.8    	 \\
 $pp \to X \to ZZ^1$    						& $5.0^{+4.3}_{-3.4}$ 	&  11.3   \\
 $pp \to X \to WH$      						& $4.5^{+5.2}_{-4.0}$   & 15.5   	 \\
 $pp \to X \to jj$      						& $91^{+53}_{-45}$   	& 170    	\\
 $pp \to X \to tb$  							& $0^{+11}_{-0}$   	& 38    	 \\
 $pp \to X \to tb$ (without ATLAS $bb\ell\nu$~\cite{Aad:2014xea})  	& $0^{+39}_{-0}$   	& 60    	 \\
 \hline
\end{tabular*}
\caption{Fitted cross sections to various final states found by combining all relevant channels. For 
the $tb$ final state we present the results both with and without the ATLAS leptonic search \cite{Aad:2014xea}
since this study contains a large under-fluctuation that significantly alters the result of our final 
fit. $^1$\emph{The $WZ$ and $ZZ$ channels contain a significant overlap in signal regions and should not be
considered as independent measurements.}}
\label{tab:xsec_fit}
\end{table}

By comparing the various fitted cross sections we can try to provide guidance on the kind of model required to
fit the current data. We also note that this is not completely speculative since a combined analysis of the above
searches finds that the standard model has a 2.9\,$\sigma$ discrepancy with the data due to various excesses present.

We first note that the required cross sections to correctly fit the data are very similar for $WZ$ and $ZZ$ (which 
can be considered as roughly the same measurement) and $WH$, see Tab.~\ref{tab:xsec_fit}. In our opinion this seriously motivates 
a model in which the resonant production particle carries charge, since the same particle can then be
responsible for both final states. In addition, the fact that $\sigma_{WZ}\sim\sigma_{WH}$ suggests to us that 
a model that predicts an equal branching ratio to these two modes should be considered.

In comparison, the dijet cross section for the signal is over an order of magnitude larger. In terms of
finding a model to fit these excesses this is convenient, since the easiest way to 
produce such a high mass resonance will be through a quark coupling of appropriate strength. The same
coupling will then automatically lead to a decay into a dijet final state.

Combining the $WZ$, $WH$, and dijet final states thus naturally leads to a model with a charged 
resonance. Working with the principle of simplicity, we may expect the couplings of this resonance
to be flavor diagonal and therefore predict that $\sigma_{tb}\sim \frac{1}{2}\sigma_{jj}$. Unfortunately there is
no evidence for a signal in the $tb$ final state and the 1\,$\sigma$ preferred region only extends to $\sim 10$~fb, which
is roughly $\frac{1}{10}\sigma_{jj}$. However, we refer the reader to the discussion in Sec.~\ref{sec:tb_searches}
where we note that the ATLAS leptonic search finds a cross section that is systematically 2\,$\sigma$ below the background
prediction. Consequently, the fit is heavily influenced by this single result and we believe that it is also wise 
to study the cross section measurement when this analysis is removed. In this case, the 1\,$\sigma$ preferred 
region extends up to $\sim 40$~fb and is therefore perfectly compatible with the observed dijet signal.

\section{Interpretation in the Left-Right Symmetric Model} 
\label{sec:LRM}

We now turn towards a specific model and interpret the observed excesses in the Left-Right Symmetric Model~\cite{Mohapatra:1986uf}. 
In Sec.~\ref{sec:LRM_intro}
we summarize the structure and key phenomenological properties of this framework. We then compare its predictions to the
experimental data and perform a model parameter fit in Sec.~\ref{sec:LRM_fit}. Finally we discuss the prospect of observing
signatures for this model
during the upcoming LHC Run~II in Sec.~\ref{sec:run2}.

\subsection{Model essentials}
\label{sec:LRM_intro}

In the Left-Right Symmetric Model (LRM) framework~\cite{Mohapatra:1986uf}, the SM
electroweak gauge group is extended to $SU(2)_L \times SU(2)_R \times U(1)'$ with the usual left-handed (right-handed) fermions of the SM
transforming as doublets under $SU(2)_{L(R)}$.
For instance, the left-handed (LH) quarks transform as $(2,1,1/3)$, whereas the right-handed (RH) quarks transform as (1,2,1/3).
The corresponding gauge couplings are denoted by $g_{L,R}$, which 
are not in general equal, as well as $g'$. As a consequence of this $L-R$ symmetry, a RH neutrino must necessarily be
introduced for each generation; neutrinos are thus naturally massive in this framework.

The SM is recovered by the breaking $SU(2)_R \times U(1)' \rightarrow U(1)_Y$,  where additional Higgs fields are required to break this symmetry
generating the masses of the new gauge fields $W_R^\pm$, $Z_R$.
This breaking relates the $g_R$ and $g'$ couplings to the usual SM 
hypercharge coupling $g_Y$. Note that since $g_L$ and $g_Y$ are both well-measured quantities, only the ratio $\kappa=g_R/g_L$
remains as a free parameter in the model. The new Higgs fields responsible for $SU(2)_R$ 
breaking are usually assumed to transform either as a doublet, $(1,2,-1)$, or as a triplet, $(1,3,-2)$, under the $SU(2)_R$ gauge group, and 
the neutral component obtains a multi-TeV scale vev, $v_R$. The corresponding LH $SU(2)_L$  doublet or triplet Higgs fields, 
$(2,1,-1)$ or $(3,1,-2)$,
which must also be present to maintain the $L-R$ symmetry, are assumed to obtain a vanishing or a tiny, phenomenologically irrelevant vev,
which we will set to zero (i.\,e.\ $v_L=0$) below.

The immediate impact of the choice of doublet versus triplet Higgs breaking is two-fold. In the triplet case, the RH 
neutrinos $N_{e,\mu,\tau}$ can obtain TeV-scale Majorana masses through the triplet vev, and thus see-saw-suppressed masses 
can be generated for the familiar LH neutrinos, which are now themselves necessarily of a Majorana character. Naively, we might 
expect the $W_R$ and $N_\ell$ masses to be of a similar magnitude. Furthermore, when triplet breaking is chosen we can 
easily identify $U(1)'$ with $U(1)_{B-L}$. If, on the other hand, doublet breaking is assumed, then the LH and RH neutrinos  
must necessarily pair up to instead form Dirac fields.

In addition to the many low-energy implications of this choice (such as potentially observable neutrinoless double beta decay
in the Majorana neutrino case, or RH leptonic currents appearing in 
$\mu$ decay in the Dirac case), this selection has an immediate impact at colliders that is relevant for our analysis. If the 
neutrinos are Dirac, then the decay $W_R^\pm \to \ell^\pm \nu_l$ occurs where the $\nu_\ell$ appear as missing energy.
The LHC Run I searches for this mode constrain the mass of $W_R^\pm$ to lie beyond the range of interest for the present analysis, even 
for very small values of $\kappa \simeq 0.15$ (which lie outside the LRM physical region as we will discuss below).
This highly disfavors the doublet breaking
scenario from our perspective. However, in the triplet breaking case, we instead find the leptonic decay now is of 
the form $W_R^\pm \to \ell^\pm N_\ell$, so that the search reach depends on the relative ordering of the $W_R$ and $N_\ell$ masses. If the 
mass relation $m_{N_\ell}>m_{W_R}$ is satisfied, then the $W_R$ has no on-shell, two-body, leptonic decay modes and these LHC Run I search constraints 
are trivially avoided. If the RH neutrinos are below the $W_R$ in mass, then $\ell \ell jj$ final states will be produced.
CMS~\cite {Khachatryan:2014dka} has observed a potential excess in this mode for the case $\ell=e$ but not for $\ell=\mu$. 
To interpret this as a real signal in the present scenario would require the various $N_\ell$ to be non-degenerate, so that 
$N_e$ ($N_\mu$) is lighter (heavier) than $W_R$~\cite{Deppisch:2014qpa,Deppisch:2014zta,Heikinheimo:2014tba,Dobrescu:2015qna,Gluza:2015goa}. 
This, however, would naively lead to a predicted branching fraction for the process $\mu \to e \gamma$~{\cite{iandl}}
which is far larger than the current experimental bound~{\cite {mu2e}}, unless the flavor and mass eigenstates of the $N_\ell$
are extremely well-aligned. 
Furthermore, since the $N_\ell$ are Majorana states, their decays produce the final states of like-sign as well as opposite sign leptons
with equal 
probability, which is not what CMS apparently observes. To avoid these issues here we will thus assume both 
triplet $SU(2)_R$ breaking and that the relation $m_{N_\ell} > m_{W_R}$ is satisfied. 

Going further, we note that since the LH (RH) SM fermions transform as doublets under $SU(2)_{L(R)}$, their masses must be 
generated by the introduction of one or more bi-doublet scalars, i.\,e.\ fields transforming as doublets under both 
$SU(2)_{L(R)}$ groups simultaneously as $(2,2,0)$. The vevs of these bi-doublets (with each bi-doublet, $\Phi_i$, having 
two distinct vevs, $k_{1,2i}$ which are of order of the electroweak scale) break the SM gauge group in the usual manner and act similarly to 
those that occur in Two-Higgs Doublet Models~\cite{Branco:2011iw}. Given a sufficiently extensive set of bi-doublets, it is 
possible to construct models wherein the CKM matrices in the LH and RH sectors are uncorrelated. However, the relationship 
$|V_{ij}^L| = |V_{ij}^R|$ is the more conventional result if we want to avoid flavor-changing neutral currents; we will assume the validity of this 
relationship in the analysis below to greatly simplify the discussion as this additional parameter freedom is not needed 
here to explain the data. 

We can now write the full $W-W_R$ mass matrix in a generic manner as follows:
\begin{eqnarray}
{\cal M_W}^2 & =  \left( \begin{array}{cc}
                         m_W^2 & \beta_w m_W^2  \\
                         \beta_w m_W^2 & m_{W_R}^2 
                         \end{array}\right) \,.
\end{eqnarray}
Here one finds that $m_W^2= {g_L^2} \sum_i (k_{1i}^2+k_{2i}^2)/4$, which we note is the would-be SM $W$ mass,  
and correspondingly $m_{W_R}^2= {g_R^2} (2v_R^2 +\sum_i (k_{1i}^2+k_{2i}^2))/4$. Note that, apart from a $\mathcal{O}(1)$ coefficient $\beta_w$,
the off-diagonal term is proportional to $m_W^2$. The reason for this is that the 
off-diagonal terms in the mass matrix are also generated by the vevs $k_{1,2i}$, so they are naturally of the order of the weak scale;
one finds explicitly that  
\begin{equation}
\beta_w =\kappa~ \frac{2\sum k_{1i}k_{2i}}{\sum_i (k_{1i}^2+k_{2i}^2)}\,.     
\end{equation}
To diagonalize this matrix we rotate the original $W, W_R$ fields into the mass eigenstates $W_{1,2}$ (where $W_1$ is 
identified as the well-known lighter state) via a mixing angle $\phi_w$ given by 
\begin{equation}
\tan 2\phi_w ={\frac{-2\beta_w m_W^2}{m_{W_R}^2-m_W^2}}\,.
\end{equation}
When $m_{W_R}^2 \gg m_W^2$, as in the case under consideration, we obtain that $\phi_w \simeq -\beta_w (m_W/m_{W_R})^2$. 
Note that in the most simple, single bi-doublet case, we find that $\beta_w =2\kappa \tan \beta/(1+\tan ^2 \beta)$.
Here we have defined the ratio of the $k_{1,2}$ vevs  as $\tan \beta$, as usual.
Although $W_{1,2}$ are the mass eigenstates, for clarity we will continue to refer to them as $W,W_R$.

\begin{figure}
 \centering \vspace{-0.0cm}
 \includegraphics[width=0.49\textwidth]{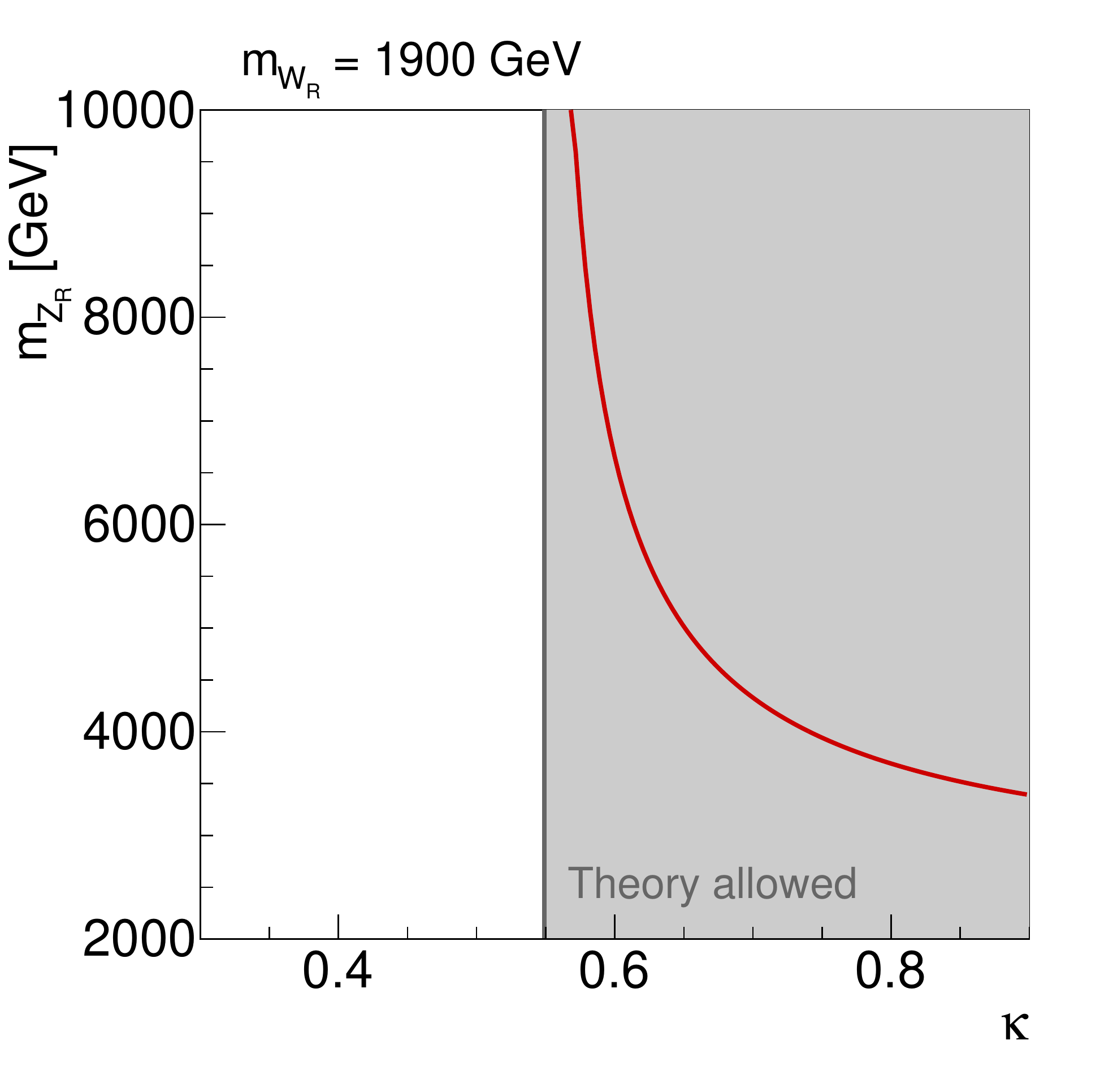}
 \caption{$m_{Z_R}$ as a function of $\kappa$ for $m_{W_R} = 1900$~GeV. The physical region is shaded in grey.}
 \label{fig:kappa_mZR}
\end{figure}

We can perform a similar analysis in the $Z-Z_R$ mixing case. This is simplified by first going to the basis where the 
massless photon is trivially decoupled, reducing the original $3 \times 3$ mass matrix to one which is only $2 \times 2$.
Then the SM $Z$ couples as usual as ${\frac{g_{L}}{c_w}}(T_{3L} -x_wQ)$, where $c_w=\cos \theta_w$, $Q$ is the electric charge, 
$x_w= s_w^2 = \sin^2 \theta_w$ and $T_{3L}$ is the 3$^{\mathrm{rd}}$ component of the LH weak 
isospin. Recalling $Q=T_{3L}+T_{3R}+(B-L)/2$
we can write the analogous $Z_R$ coupling as
\begin{equation}
    \mathcal{O}_{Z_R} = \frac{g_L}{c_w}[\kappa^2-(1+\kappa^2)x_w]^{-1/2}[x_wT_{3L}+\kappa^2(1-x_w)T_{3R}-x_wQ]\,. \label{eq:O_Z_R}
\end{equation}
Interestingly, since we know that $v_R^2 \gg k_{1,2i}^2$, the mass ratio of the \emph{physical} $W_R$ and $Z_R$ is given to a 
very good approximation by simply setting the $k_{1,2i}^2 \to 0$, i.\,e.\ 
\begin{equation}
{\frac{m_{Z_R}^2}{m_{W_R}^2}}={\frac{\kappa^2(1-x_w)\rho_R}{\kappa^2(1-x_w)-x_w}}>1\,,
\label{eq:mZprime}
\end{equation}
with the values of $\rho_R=1 (2)$ depending upon whether $SU(2)_R$ is broken by either Higgs doublets (or by triplets); in this work 
$\rho_R=2$ follows from our assumption of triplet breaking. We demonstrate this relation between $\kappa$ and the physical masses
in Fig.~\ref{fig:kappa_mZR}.

In further analogy with the $W-W_R$ case we find that the $Z-Z_R$ 
mixing angle is given by $\phi_z \simeq -\beta_z (m_Z/m_{Z_R})^2$ when $Z_R$ is heavy. $\beta_z$ is again 
an $\mathcal{O}(1)$ parameter which is generally a ratio of the various bi-doublet vevs. In the case of a single bi-doublet, the 
general expression for $\beta_z$ simplifies significantly to $\beta_z=-[\kappa^2-(1+\kappa^2)x_w]^{1/2}$. Note that for $\kappa$ 
values near the theoretical minimum (as we will discuss next) we find that $\beta_z$ is very small, implying a further 
suppression of $Z-Z_R$ mixing in this case.

Examining the expressions for both the $Z_R/W_R$ mass ratio, as well as that 
for the $Z_R$ couplings, we see that $\kappa > [x_w/(1-x_w)]^{1/2} \simeq 0.55$ is required for the fields to remain physical 
as alluded to in the discussion above. Below this value the $Z_R$ coupling becomes imaginary, see Eq.~\eqref{eq:O_Z_R}, and 
$m_{Z_R}^2$ is negative, c.\,f.\ Eq.~\eqref{eq:mZprime}. This theoretical requirement will play an important role in the discussion of our fit results below.

In this scenario, the partial width of the $W_R$ into quark pairs is given by~\cite{Rizzo:1981dm,Rizzo:1981su}
\begin{align}
  \label{eq:BRjj}
  \Gamma(W^+_R \to u \bar{d}) &= \Gamma(W^+_R \to c \bar{s}) = 3 \kappa^2 A \left(1 + \frac {\alpha_s (m_{W_R})} \pi \right) \,, \\
  \Gamma(W^+_R \to t \bar{b}) &= 3 \kappa^2 A \left(1 + \frac {\alpha_s (m_{W_R})} \pi \right) \left (1- \frac {m_t^2} {m_{W_R}^2} \right)^2 \left(1+\frac 1 2 \, \frac {m_t^2} {m_{W_R}^2}\right)\,,
  \label{eq:BRtb}
\end{align}
where $A = G_F m_W^2 m_{W_R} / (6 \pi \sqrt{2})$ is an overall constant.

Calculating the decays into diboson states is a little more complicated since correctly
obtaining the effective $W_R WZ$ coupling in the LRM is subtle.  As in the SM, the trilinear couplings of the gauge bosons 
arise from the non-abelian parts of the kinetic terms for the gauge fields, in particular, from the part 
of the covariant derivative containing the gauge fields acting on themselves. In the basis where the massless photon 
explicitly appears, the covariant derivative is given by   
\begin{equation}
{\cal D}=\partial -ieQA- \frac{i}{\sqrt 2} g_L T_L^\pm \cdot W^\mp + (L\to R)-i{\frac{g_L}{c_w}}(T_{3L}-x_wQ)Z-i \mathcal{O}_{Z_R}Z_R\,,
\end{equation}
where we have suppressed the Lorentz index and where the $Z_R$ coupling operator $\mathcal{O}_{Z_R}$ is given in Eq.~\eqref{eq:O_Z_R}. $\cal D$ 
acting on the $W,W_R$ generates both a $W^+W^-Z$ coupling, as in the SM, as well as a $W_R^+W_R^-Z$ coupling. Since  
$W\,,W_R$ have $Q=1$ and $T_{3L}(W,W_R)=1,0$, these two couplings are simply ${\frac{g_L}{c_w}}(1-x_w)$and 
${\frac{g_L}{c_w}}(-x_w)$, respectively. In terms of the mass eigenstates $W=W_1 c +W_2 s$ and $W_R= W_2 c-W_1 s$ (where 
$s=\sin \phi_w, c=\cos \phi_w$), corresponding to the physical masses $m_{1,2}$, the off-diagonal $W_1^\pm W_2^\mp Z$ coupling 
can be obtained by combining these two individual contributions. We obtain 
\begin{equation}
 {\frac{g_L}{c_w}}[cs(1-x_w)+(-sc)(-x_w)] = {\frac{cs \, g_L}{c_w}}\,.
\end{equation}
This reproduces the result obtained some years ago~\cite{Deshpande:1988qq}.

In the expressions above, we have ignored any $Z-Z_R$ mixing since it is numerically small in the parameter range 
of interest in our scenario since both $|\beta_z| \ll |\beta_w|$ and $m_{W_R}^2/m_{Z_R}^2 \ll 1$. However, we note that by 
identifying $Z$ with $Z_1$, an additional contribution will arise from the $W_R^+W_R^-Z_R$ coupling (the 
corresponding $W^+W^-Z_R$ coupling is absent as can be seen from the structure of the operator $O_{Z_R}$), but this interaction is 
relatively suppressed by an additional factor of the $Z-Z_R$ mixing angle, $\phi_z$, so we will ignore this term in our analysis.  

In the corresponding partial width for the decay $W_R\to WZ$ (i.\,e.\ for $W_2\to W_1 Z_1$) the above coupling will appear  
quadratically and is always accompanied by an additional factor of $m_2^4/(m_1^2m_Z^2)$ arising from the longitudinal parts 
of the corresponding gauge boson polarization appearing in the final state.  Now, 
since $\phi_w \simeq \beta_w m_W^2/m_{W_R}^2$ and $m_{1(2)}^2=m_{W,W_R}^2[1+\mathcal{O}(m_W^2/m_{W_R}^2)]$, using the 
SM relation $m_W=c_w m_Z$, we see that the large mass ratios will  cancel, leaving us with just an overall dependence of 
$\sim g_L^2 \beta_w^2[1+\mathcal{O}(m_W^2/m_{W_R}^2)]$. In the single bi-doublet 
model this reduces further to $\sim g_R^2 \sin^2 2\beta ~[1+\mathcal{O}(m_W^2/m_{W_R}^2)]$ with $\tan \beta$ being the ratio of the two 
bi-doublet vevs as defined above. 

As shown in explicit detail recently in Ref.~\cite{Dobrescu:2015yba} (which we have 
verified), the corresponding square of the $W_R WH$ coupling, where $H$ is to be identified with the (almost) SM Higgs, is 
given by $\sim g_R^2 \cos^2 (\alpha+\beta)$, where $\alpha$ is the mixing angle of 
the two Higgs doublet model \cite{Branco:2011iw}. This is at the same level of approximation where higher order terms in the gauge boson  
mixings are neglected. Going to the Higgs alignment limit, i.\,e.\  $\alpha \simeq \beta-\pi/2$, to recover the SM-like Higgs,  
one observes that $\cos^2 (\alpha+\beta) \to \sin^2 2\beta$. This demonstrates the equality of the effective $W_RWH$ and 
$W_RWZ$ couplings up to higher order terms in the various mass ratios, as is required by the Goldstone boson equivalence theorem.  

To cut a long story short, the partial decay widths of the $W_R$ into diboson states are given by
\begin{align}
  \Gamma(W^+_R \to W^+ Z) &= \frac {A } 4 \, a_w^2 
\left( 1 - 2 \frac{m_W^2 + m_Z^2} {m_{W_R}^2} + \frac {(m_W^2 - m_Z^2)^2} {m_{W_R}^4} \right)^{3/2} \notag \\
&\phantom{=} \quad \times \left(1 + 10 \frac{m_W^2 + m_Z^2} {m_{W_R}^2} + \frac {m_W^4 + 10 m_W^2 m_Z^2 + m_Z^4} {m_{W_R}^4} \right) \,, 
  \label{eq:BRWZ}\\
  \Gamma(W^+_R \to W^+ H) &= \frac {A } 4 \, a_H^2 
   \left( 1 - 2 \frac{m_W^2 + m_H^2} {m_{W_R}^2} + \frac {(m_W^2 - m_H^2)^2} {m_{W_R}^4} \right)^{1/2} \notag \\
&\phantom{=} \quad \times \left(1 + \frac{10m_W^2 - 2m_H^2} {m_{W_R}^2} + \frac {(m_W^2 - m_H^2)^2} {m_{W_R}^4} \right)
  \label{eq:BRWH}
\end{align}
with $a_w = (cs) (m_{W_R}^2 / m_W^2)$. $a_H$ depends on the details of the Higgs sector; however,
the equivalence theorem requires that $a_H = a_w + \mathcal{O} (m_W^2 / m_{W_R}^2).$

\subsection{Is that it? Fitting the Left-Right Symmetric Model to data}
\label{sec:LRM_fit}

\begin{figure}
 \centering \vspace{-0.0cm}
 \includegraphics[width=0.49\textwidth]{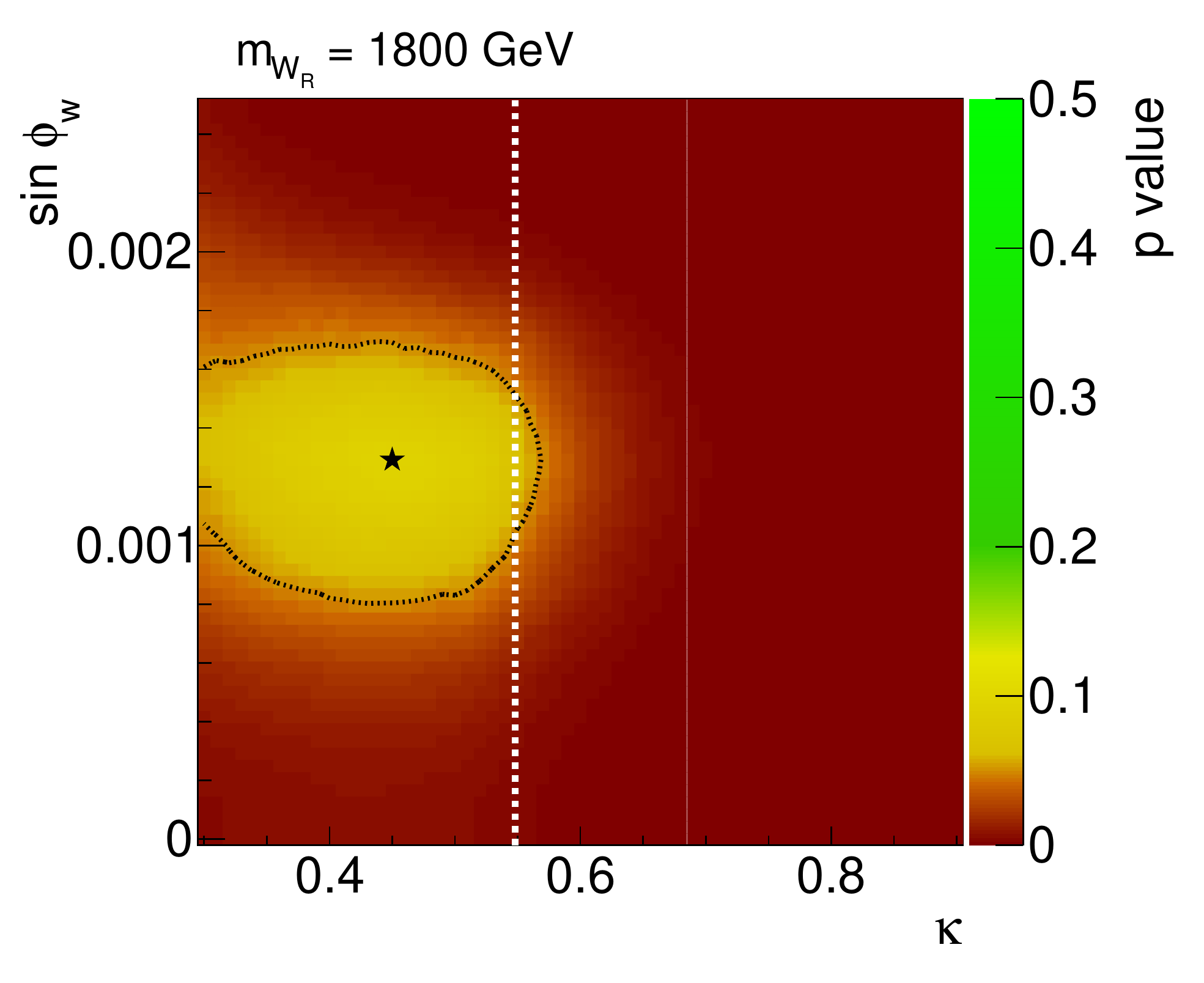}
 \includegraphics[width=0.49\textwidth]{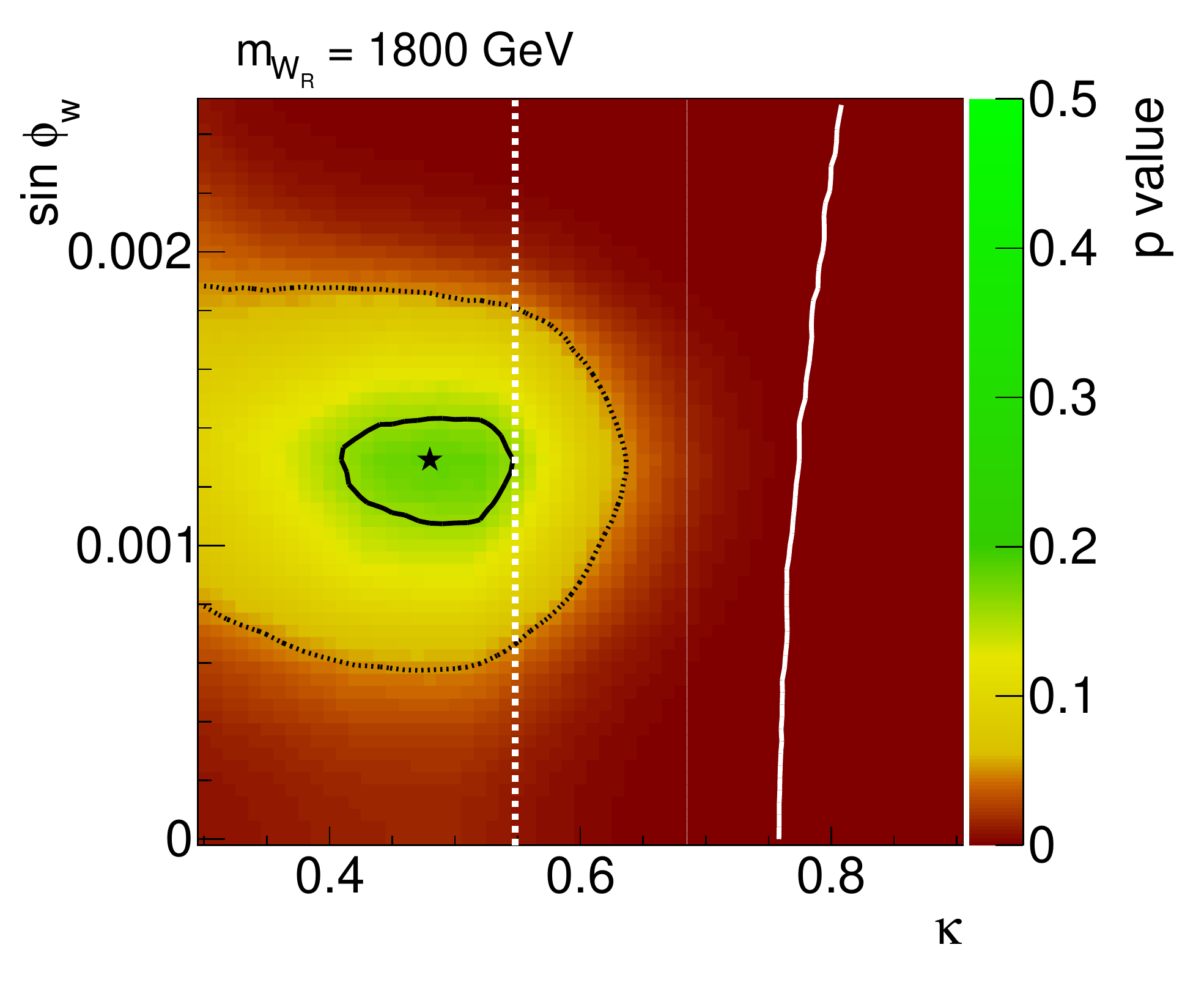}
 \caption{Fit of the LRM parameters $\kappa$ and $\sin \phi_W$ to the searches listed 
 in Tab.~\ref{tab:xsec_fit} with $m_{{W_R}}=1800$~GeV. The region compatible with data at
 68\% (90\%) CL is shown with a black solid (dotted) line. The white dashed line denotes
 the theoretical limit $\kappa > s_W / c_W \simeq 0.55$.  The star represents the best fit point. Left: fit including the ATLAS
 leptonic $tb$ search that systematically disfavors the full parameter space as well as the
 background-only hypothesis with $\sim$1.8\,$\sigma$. Right: 
 fit excluding the ATLAS leptonic $tb$ search. The white solid line shows the $95 \%$~CLs limit
 from the ATLAS leptonic $tb$ search (the region left of the line is allowed).}
 \label{fig:kappa_fit_1800}
\end{figure}

\begin{figure}
 \centering \vspace{-0.0cm}
 \includegraphics[width=0.49\textwidth]{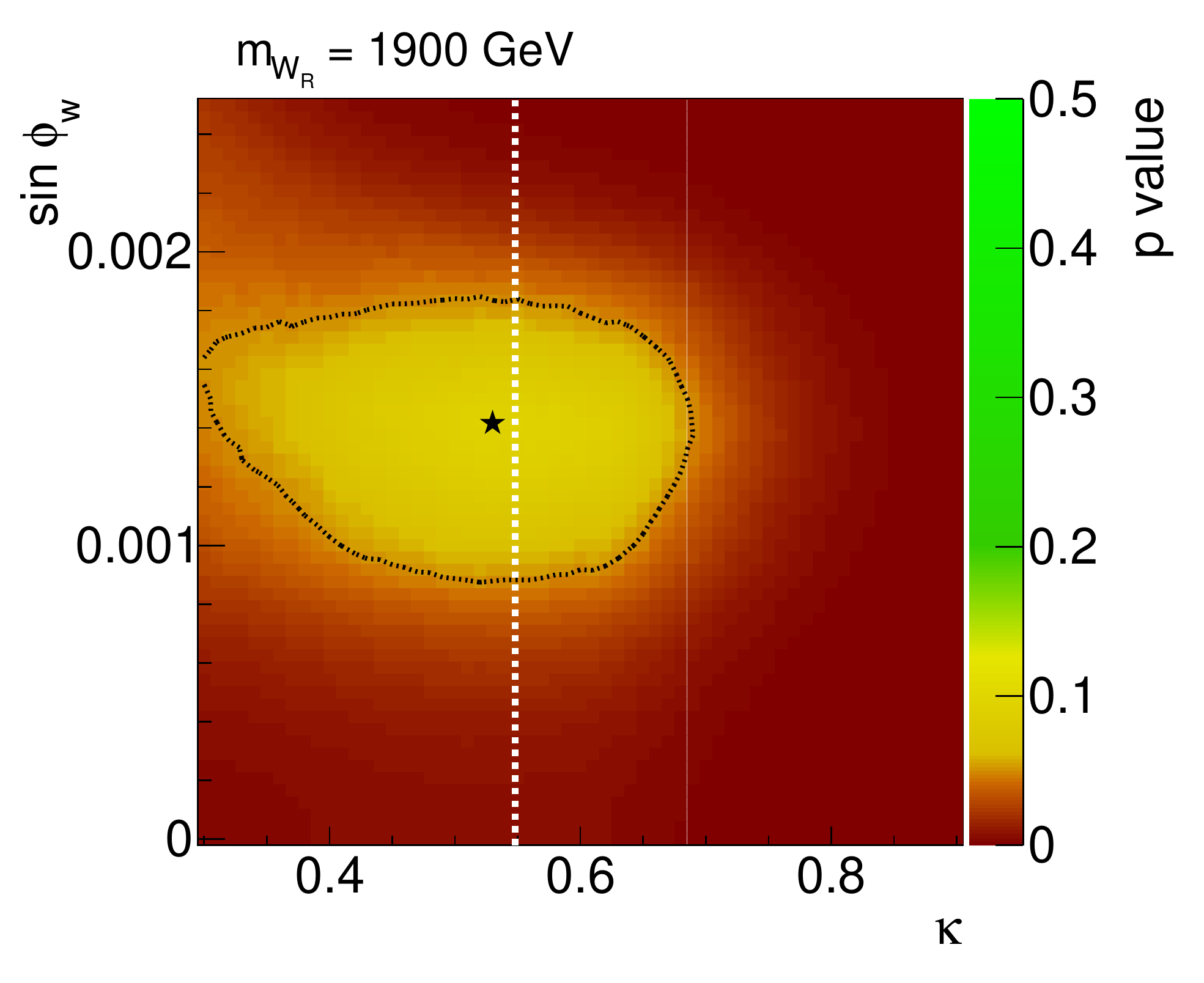}
 \includegraphics[width=0.49\textwidth]{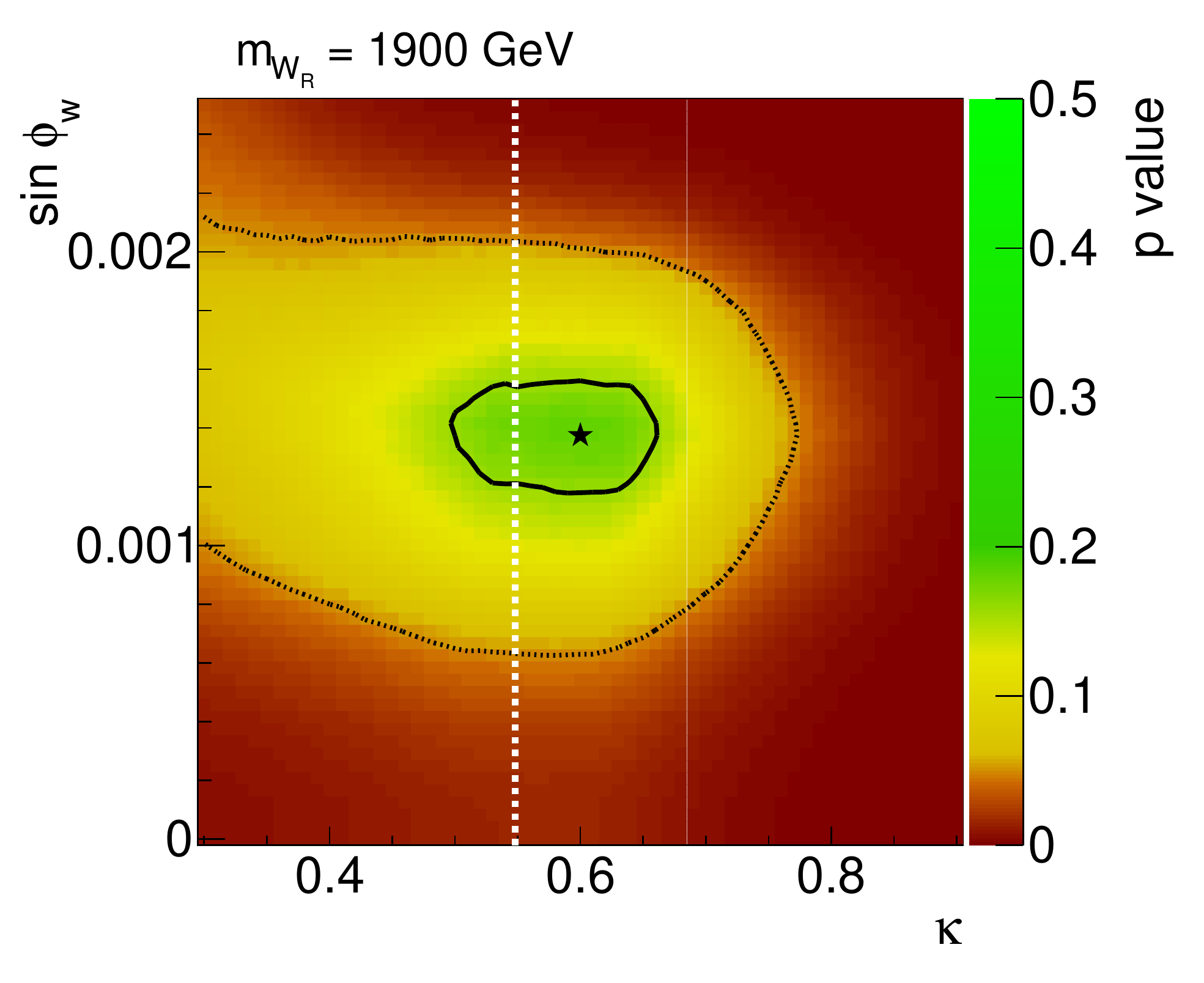}
 \caption{Fit of the LRM parameters $\kappa$ and $\sin \phi_W$ to the searches listed 
 in Tab.~\ref{tab:xsec_fit} with $m_{{W_R}}=1900$~GeV. The region compatible with data at
 68\% (90\%) CL is shown with a black solid (dotted) line. The white dashed line denotes
 the theoretical limit $\kappa > s_W / c_W \simeq 0.55$.  The star represents the best fit point.  Left: fit including the ATLAS
 leptonic $tb$ search that systematically disfavors the full parameter space as well as the
 background-only hypothesis with $\sim$1.8\,$\sigma$. Right: 
 fit excluding the ATLAS leptonic $tb$ search. The $95 \%$~CLs limit
 from the ATLAS leptonic $tb$ search is outside the plotted region and does not affect
 the preferred region.}
 \label{fig:kappa_fit_1900}
\end{figure}

We now investigate whether the excesses observed at the LHC can be explained by a $W_R$ resonance of the Left-Right Symmetric Model.
In section \ref{sec:measurements} we concluded that the data favors a charged resonance with approximately equal branching
ratios to $WZ$ and $WH$. This agrees very well with the predictions of the LRM, where these branching ratios are equal up to
$\mathcal{O} (m_W^2 / m_{W_R}^2)$ corrections. The excess observed in the dijet channels, together with the not so restrictive bounds in the $tb$
searches, are also promising.

In such an interpretation, $m_{W_R}$ should take on the value 1800\,--\,1900~GeV to be compatible with the observed
excesses. A mass of 2000~GeV, as favoured by the ATLAS diboson excess alone, appears to be disfavoured by constraints from the
other channels, such as the semileptonic diboson searches. We will perform fits for the two scenarios $m_{W_R} = 1800, 1900 \ \GeV$, in the understanding
that the limited number of events, width effects and experimental resolution makes it difficult to pin down the mass more precisely.

In order to test the compatibility of the LRM with the data, we include the parameter space of the LRM in our cross-section fit.  Again, we
first calculate the compatibility of a parameter point with the observed event numbers in all individual analyses.
This is followed by a combination of the results, giving an overall $p$ value which we present the results in terms of best fit points and
confidence regions at $68 \%$ and $90\%$~CL in the parameter space.

The narrow width approximation is used throughout. We find that the width of the $W_R$ is of order 1\,--\,2$\%$ of its mass in the best-fit
region, making the error due to this approximation sub-dominant to the other uncertainties 
present. We calculate the production cross section of the $W_R$
using the MMHT2014 NNLO pdfs~\cite{Harland-Lang:2014zoa} with constant NNLO $k$~factors~\cite{Melnikov:2006kv} while
the branching ratios are calculated using Eq.~\eqref{eq:BRjj}\,--\,\eqref{eq:BRWH}.
We do not assume $a_H = a_w$, but allow for corrections of order $\mathcal{O} (m_W^2 / m_{W_R}^2)$ by introducing a  parameter $\xi$ using
\begin{equation}
  a_H = a_w + \xi \frac {m_W^2} {m_{W_R}^2} \,.
\end{equation}
$\xi$ is allowed to float in the range $[-10, 10]$ with a flat prior and is profiled
over in our results.

\begin{figure}
 \centering \vspace{-0.0cm}
 \includegraphics[width=0.49\textwidth]{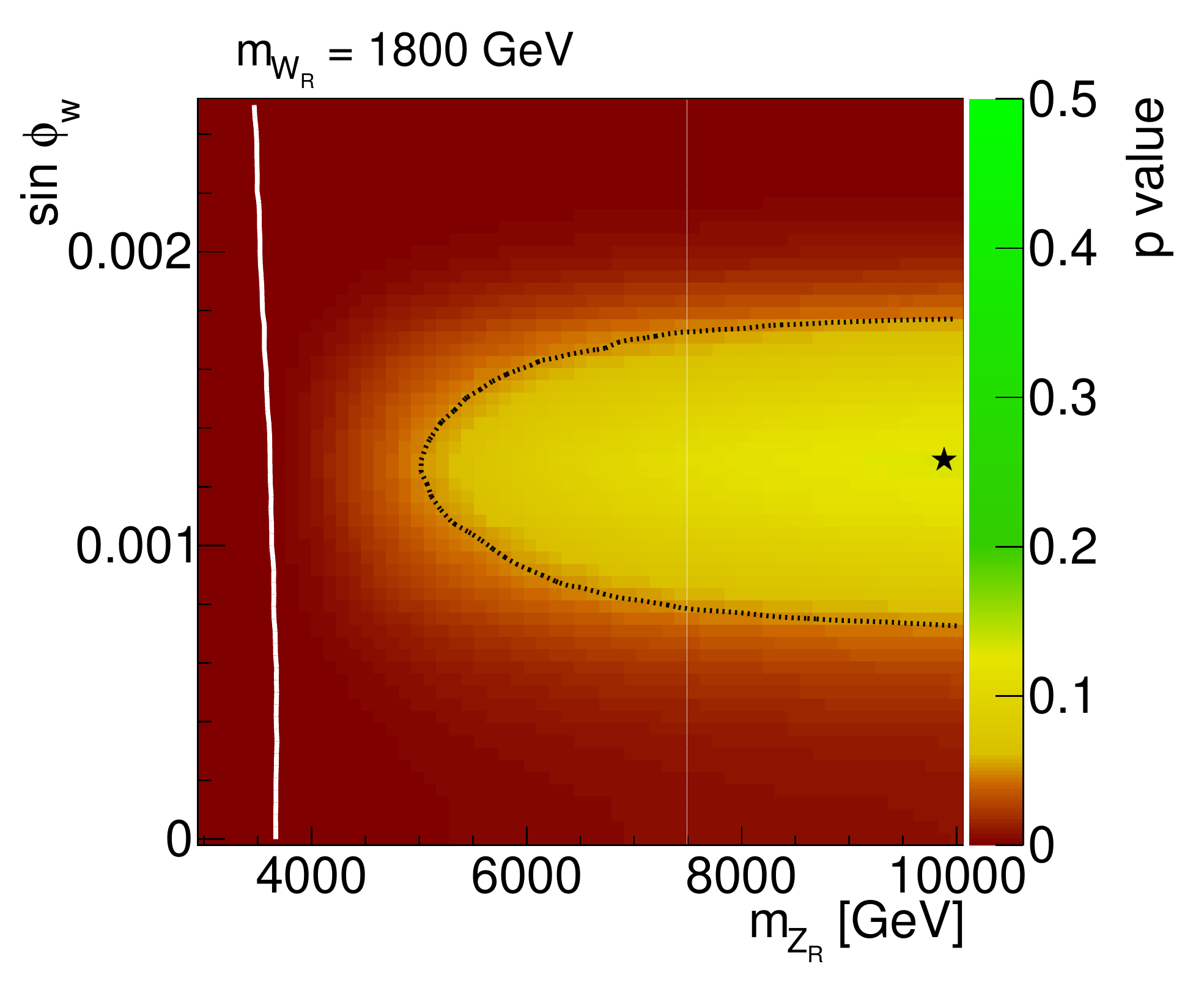}
 \includegraphics[width=0.49\textwidth]{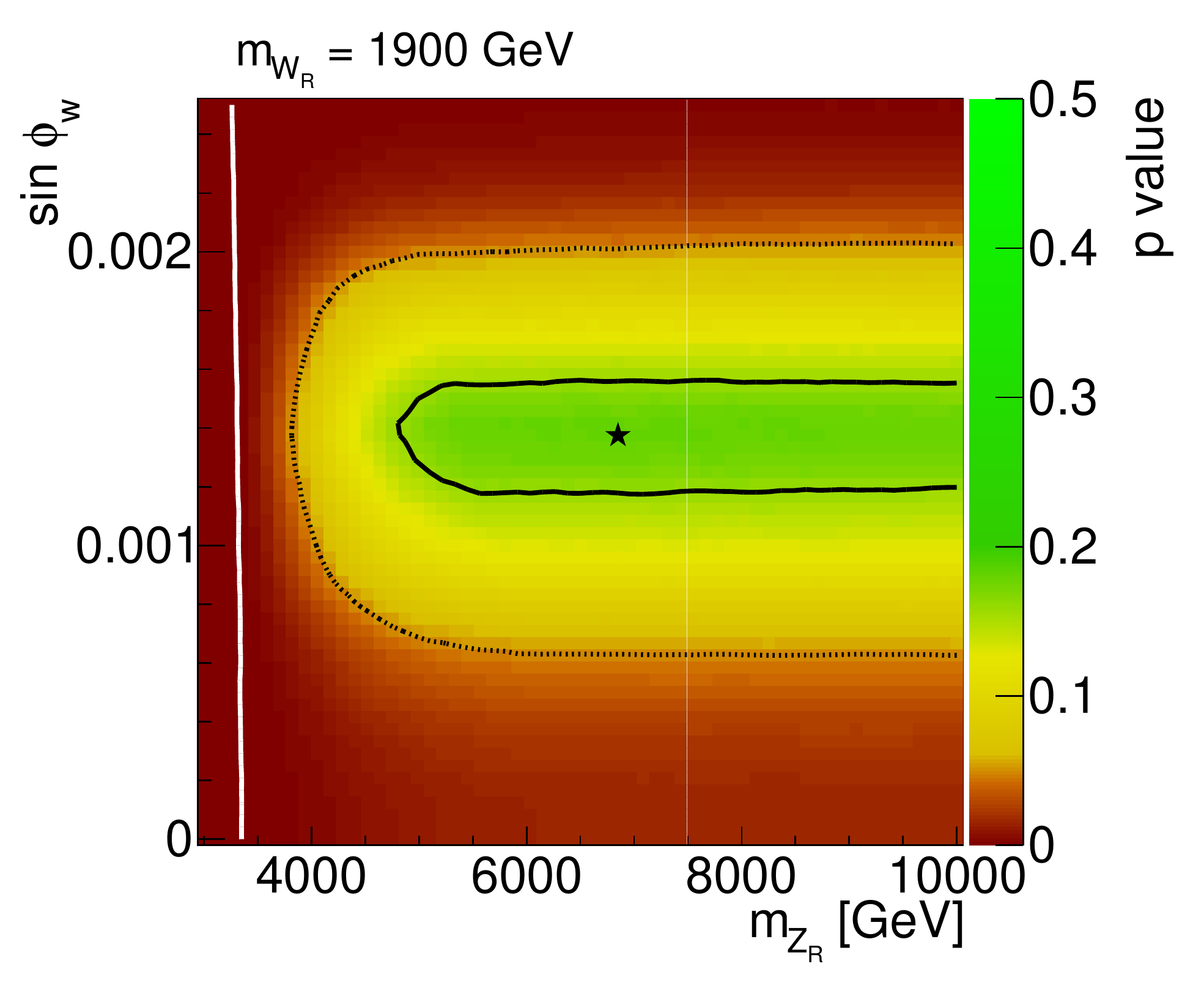}
 \caption{Fit of the LRM parameters $m_{Z_R}$ and $\sin \phi_W$ to the searches listed 
 in Tab.~\ref{tab:xsec_fit} with $m_{{W_R}}=1800$~GeV (left) and $m_{{W_R}}=1900$~GeV (right). The region compatible with data at
 68\% (90\%) CL is shown with a black solid (dotted) line. The star represents the best fit point. We exclude the ATLAS leptonic $tb$ search from the fit, but the white solid line shows the $95 \%$~CLs limit
 from this search (the regions right of the lines are allowed).}
 \label{fig:mZprime_fit}
\end{figure}

In Fig.~\ref{fig:kappa_fit_1800} and \ref{fig:kappa_fit_1900} we present the results from our fit for $m_{W_R} = 1800~\GeV$ and
$m_{W_R} = 1900~\GeV$ respectively. In the left panels
we show the overall agreement with data when all experimental studies presented
in the previous section are included. In this case, no part of the parameter space of the LRM is compatible with data at $68 \%$~CL.
This is due to the tension between the dijet excess in the ATLAS and CMS searches and the ATLAS $tb$ search in the leptonic decay mode.
However as we have argued in Sec.~\ref{sec:vec_higgs}, this ATLAS $tb$ search finds a rate consistently $2 \,\sigma$ below the background expectation, and
while this does not lead to a strong CLs limit, it severely punishes the overall $p$ value in our fit for the full parameter space.

We therefore also present results where this single ATLAS $tb$ search is excluded from the fit itself, where we explicitly check that the best-fit regions
are not excluded by the CLs limits from this study. The results are shown in the right panels of Fig.~\ref{fig:kappa_fit_1800}
and \ref{fig:kappa_fit_1900}. There is now a well-defined region where the LRM agrees very well with all searches and can describe 
the observed excess while satisfying the constraints from the other searches. For $m_{W_R} = 1800~\GeV$ this region is roughly given by
$0.4 \lesssim \kappa \lesssim 0.55$ and $0.00011 \lesssim \sin \phi_w \lesssim 0.0015$. For $m_{W_R} = 1900~\GeV$ the smaller production cross section
allows for larger couplings $0.5 \lesssim \kappa \lesssim 0.65$ and $0.0012 \lesssim \sin \phi_w \lesssim 0.0016$.

These preferred couplings fall into a special place in the parameter space: as described in Sec.~\ref{sec:LRM_intro}, the theory requires $\kappa > s_W / c_W \simeq 0.55$ to be
consistent. This means that for $m_{W_R}=1800~\GeV$ the preferred region at $1\,\sigma$ falls entirely in the unphysical regime.
The situation is different in the $m_{W_R} = 1900~\GeV$ scenario, where we find good agreement further from this boundary of the physically allowed region.

Since the mass of the $Z_R$ is fixed by $\kappa$ and $m_{W_R}$ (as shown in Eq.~\eqref{eq:mZprime} and Fig.~\ref{fig:kappa_mZR}), 
we may also ask what mass
the $Z_R$ must have for the LRM to be consistent with the observations. A coupling close to the boundary $\kappa \gtrsim 0.55$ corresponds to
a very heavy $Z_R$, while large values of $\kappa$ translate to lower $Z_R$ masses. In Fig.~\ref{fig:mZprime_fit} we show the results of our fit,
still excluding the ATLAS leptonic $tb$ search, in terms of $m_{Z_R}$ and $\sin \phi_w$.

\begin{figure}
 \centering \vspace{-0.0cm}
 \includegraphics[width=0.49\textwidth]{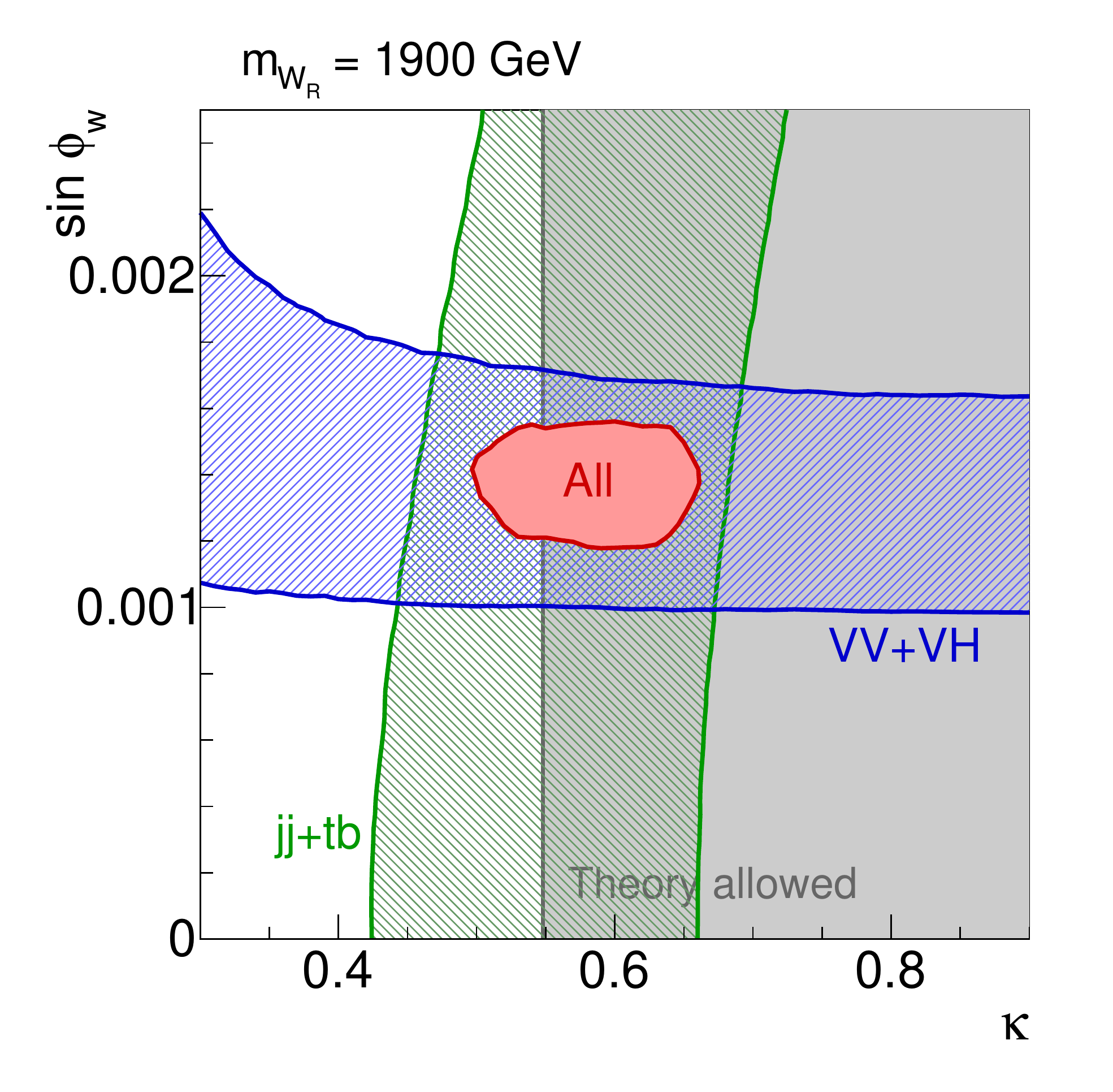}
 \includegraphics[width=0.49\textwidth]{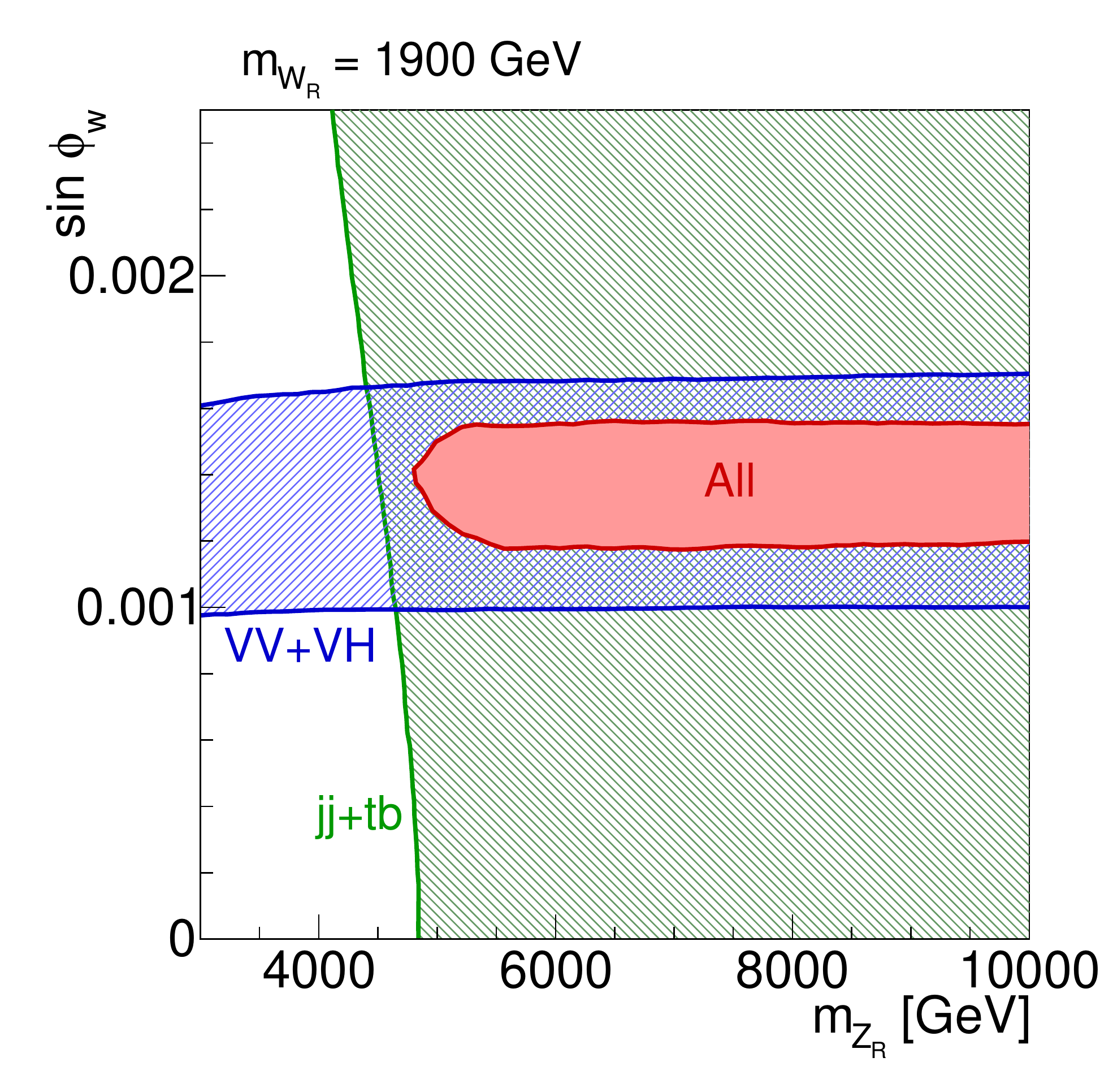}
 \caption{Preferred regions at $68\%$~CL for $m_{{W_R}}=1900$~GeV. We separately show the constraints from bosonic (blue, dark hatched) and 
 fermionic (green, light hatched) final states.
   The ATLAS leptonic $tb$ search is excluded from the fit.
   Left: parameterisation with the LRM parameters $\kappa$ and $\sin \phi_W$. The grey shaded region marks the theoretically allowed region of
   $\kappa > s_W / c_W \simeq 0.55$.
   Right: parameterisation in terms of $m_{Z_R}$ and $\sin \phi_W$.}
 \label{fig:fit_contributions}
\end{figure}

\begin{figure}
 \centering \vspace{-0.0cm}
 \includegraphics[width=0.49\textwidth]{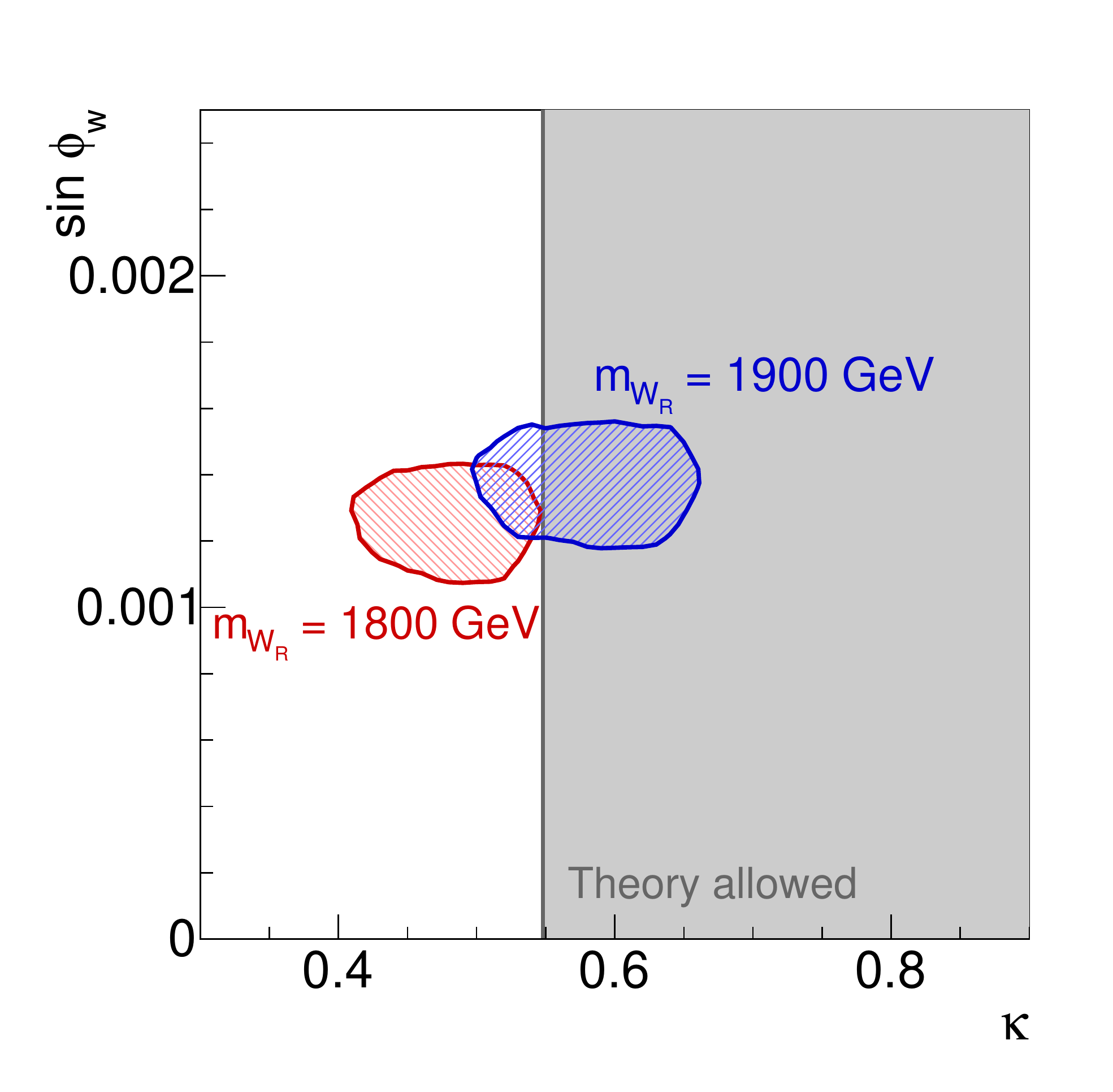}
 \caption{Preferred regions at $68\%$~CL for two different values for the $W_R$ mass. The ATLAS leptonic $tb$ search is excluded from the fit.
   The grey shaded region marks the theoretically allowed region of
   $\kappa > s_W / c_W \simeq 0.55$.}
 \label{fig:fit_masses}
\end{figure}

Assuming $m_{W_R} = 1900~\GeV$, the data permit a lower bound on $m_{Z_R}$ of around 4~\tev which
is substantially above the masses probed so far at the LHC \cite{Aad:2014cka,Khachatryan:2014fba,Patra:2015bga}, $m_{Z_R}\leq3$\tev.
There is no upper bound on $m_{Z_R}$ from our fit. This directly follows from the fact 
that our preferred regions for $\kappa$ extend into the
region $\kappa \sim 0.55$, where $m_{Z_R}$ becomes very large.

Our fit also allows us to analyze the origin of the constraints. In Fig.~\ref{fig:fit_contributions} we show the constraints from diboson and fermionic
final states separately. As expected from Eq.~\eqref{eq:BRjj} and \eqref{eq:BRtb}, the dijet and $tb$ rates fix the overall coupling constant $g_R$ while the $WZ$ and $WH$ rates 
then set the mixing angle $\phi_w$. We conclude this section with a comparison of the preferred regions for $m_{W_R} = 1800~\GeV$
and $m_{W_R} = 1900~\GeV$ in Fig.~\ref{fig:fit_masses}.

\subsection{Prospects for Run II}
\label{sec:run2}

As the LHC begins operations at 13~TeV, the prospects to discover or exclude the model presented here are excellent
due to the steep rise in heavy particle production at higher energies. In this section we first estimate
the amount of data that will be required to more thoroughly probe a possible $W_R$ resonance at
1.9~TeV. We then more speculatively explain the prospects that the LHC may produce and detect
the neutral $Z_R$ gauge boson.

At the 13 TeV LHC, the production cross-section for a $W_R$ resonance at 1.9~TeV is over 6 times higher than at 8~TeV.
Consequently, if we assume that the background scales roughly in proportion with the signal, a mere 5~fb$^{-1}$ 
will already probe the model in more detail than the current data set. Indeed, if no signal is observed, the dijet
resonance search can already be expected to exclude our best-fit point at 95\% CLs. This 
would also place the whole model under significant strain since it is the $jj$ cross-section measurement
that drives the overall coupling determination in our fit. If we do not see a continued excess here, 
the model is driven to couplings of unphysically small sizes for this value of the $W_R$ mass.

As Run II accumulates 10~fb$^{-1}$, a signal should be observed in the $tb$ final state otherwise the model
assumption of flavour diagonal couplings will start to be under significant tension. On the gauge coupling side, 
our best fit model point can easily be excluded by both $WZ$ and
$WH$ searches with less than 15~fb$^{-1}$.

A categorical 5$\sigma$ discovery does require larger data sets. Perhaps surprisingly given that most of the 
theoretical excitement has revolved around the diboson excesses, in our model we can expect the $jj$ final
state to be discovered first. Indeed, using the current search as a baseline, we expect a 5$\sigma$ discovery to 
be made with approximately 20~fb$^{-1}$ if the current best-fit point is close to reality. A discovery in the $tb$ final state 
would follow shortly afterwards with 30~fb$^{-1}$. Again the gauge boson final states require more data with
approximately 50~fb$^{-1}$  required for confirmation of the $WZ$ final state, while over 100~fb$^{-1}$
is expected to be required before $WH$ is definitively seen.

More speculative is the discovery potential of the $Z_R$ resonance since our model best fit point lies so close
to a critical theory region. As explained in Sec.~\ref{sec:LRM_intro}, model consistency requires that 
$\kappa>0.55$, and depending on the mass assumed for the $W_R$ resonance, the best fit $\kappa$ may be below 
this, see Fig.~\ref{fig:kappa_fit_1800} and \ref{fig:kappa_fit_1900}. The problem is that at as we head towards 
$\kappa\to0.55$, the mass of the $Z_R$ rapidly increases to a value far above the LHC collision energy.

\begin{figure}
 \centering \vspace{-0.0cm}
 \includegraphics[width=0.5\textwidth, angle =90]{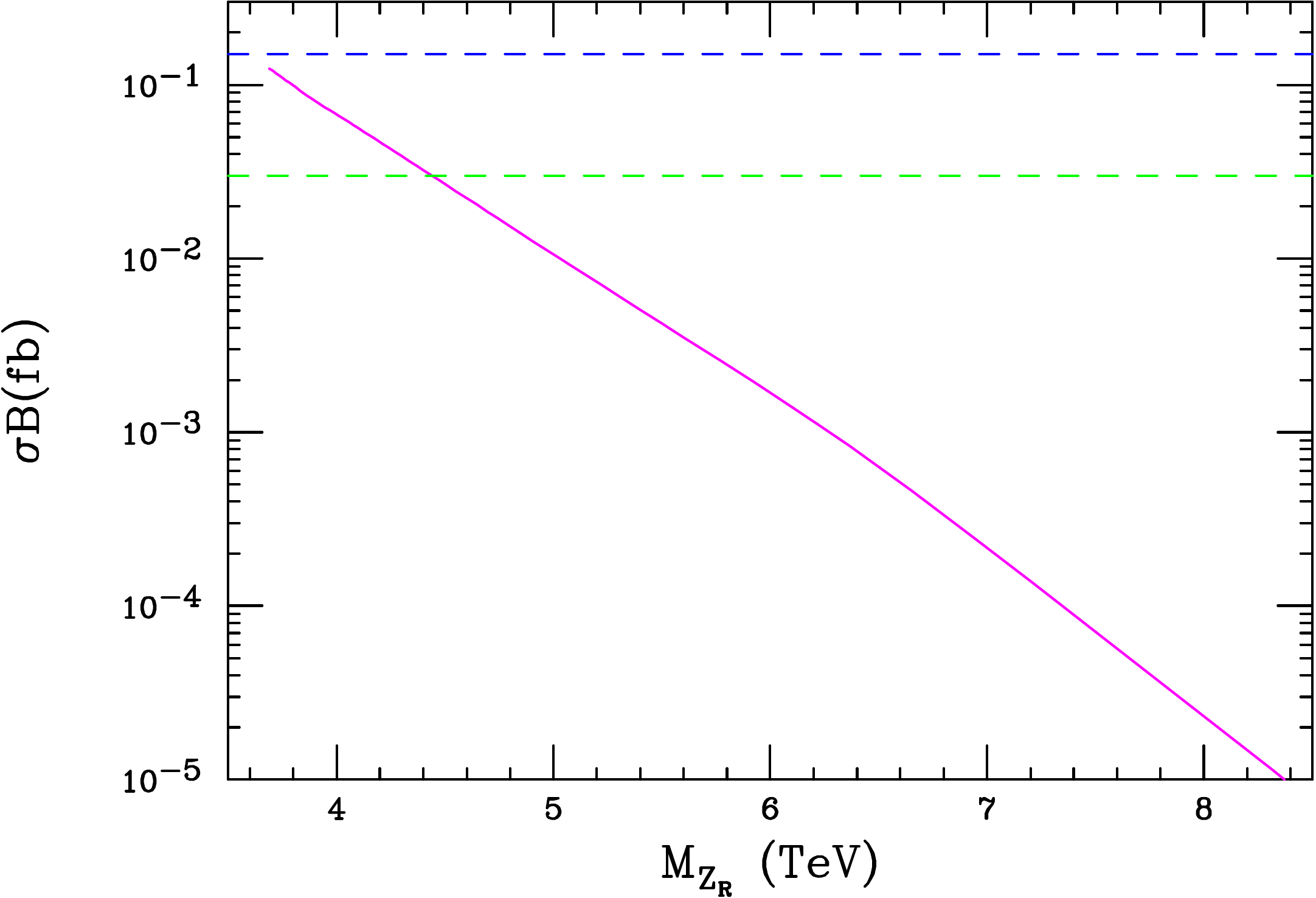}
 \caption{$Z_R$ production cross section times branching ratio into the $e^+ e^-$ and $\mu^+ \mu^-$
   states combined for $\sqrt{s} = 13$~TeV, assuming $m_{W_R} = 1900$~GeV and $0.56\leq\kappa\leq0.8$, 
   see Eq.~\eqref{eq:mZprime}. The dashed lines show the cross sections that predict
   3 expected signal events after 20 (blue) and 100 (green) fb$^{-1}$ of
   data taking. Due to the low backgrounds, these lines indicate $\sim95\%$~CLs exclusions if no signal is seen.} 
 \label{fig:ZR_13gev}
\end{figure}

Nevertheless, there is still a region of the $1\,\sigma$ preferred fit value that allows for LHC
relevant $Z_R$ masses,  especially if we assume $m_{W_R}=1900$~GeV, see 
Fig.~\ref{fig:fit_masses}. A scenario with $m_{Z_R} \gtrsim 4000$~GeV is entirely possible and in Fig.~\ref{fig:ZR_13gev}
we give an estimation of the $Z_R \to \ell^+ \ell^-$ cross section at $\sqrt{s} = 13$~TeV as a function of $m_{Z_R}$. We find that
the relevant region begins already to be probed once the LHC collects 20~fb$^{-1}$ at 13~TeV. With 100~fb$^{-1}$, the $\sim95\%$~CLs 
exclusion region stretches to $\sim 4500$~GeV.

\section{A connection to dark matter?} 
\label{sec:DM}

An immediate question that arises when hints for new physics are found at the
LHC is whether these hints could be connected to the physics of dark matter.
In the context of the anomalies discussed in this paper, such a connection
is not obvious. We will discuss four different scenarios in the following.

\textbf{a) ${W_R}$-mediated DM interactions with SM partners.}
If thermally produced DM particles $\chi^0$ are coupled to the SM sector
through a ${W_R}$ resonance, they need a charged partner $\chi^+$ which
they can annihilate into.  $\chi^+$ could be a SM lepton, in particular the
$\tau$, but in this case the DM mass would have to be smaller than $m_\tau$ to forbid
a fast DM decay via $\chi^0 \to \tau + ({W_R} \to q \bar{q})$  (the coupling of the
${W_R}$ to quarks is needed to explain the LHC anomalies discussed in the previous sections).
In the context of the Left-Right Symmetric Model, $\chi^0$ could for instance be
identified with the third generation right-handed neutrino.  Then, however,
mixing between $\chi^0$ and the other right-handed neutrinos must be forbidden
or strongly suppressed to avoid fast decays into electrons and muons.
Moreover, the ${W_R}$-mediated
annihilation cross section, which is of order
\begin{align}
  \ev{\sigma v_\text{rel}}(\chi^0\chi^0 \to \tau^+ \tau^-) \sim \frac{m_\chi^2}{M_{W_R}^4}
    \sim  10^{-30}\,\text{cm}^3/\text{sec} \,,
\end{align}
would be far below the generic value for a GeV thermal relic, $\ev{\sigma v_\text{rel}} \sim
5 \times 10^{-26}$~cm$^3$/sec~\cite{Steigman:2012nb}.  In order to obtain
the correct DM abundance in this case, significant entropy would need to
be produced in the early Universe after DM freeze-out. While this is certainly
possible, it would require a dark sector with a larger particle content and
much richer dynamics than envisioned here.  Alternatively, RH neutrino DM could
be produced via a freeze-in mechanism. A scenario along these lines has been
studied in Ref.~\cite{Frere:2006hp}.

\textbf{b) ${W_R}$-mediated DM interactions with charged partners beyond the SM.}
If both the $\chi^0$ and $\chi^+$ are new particles, a thermal freeze-out scenario
is possible if the $\chi^+$ is $\lesssim 10\%$ heavier than the $\chi^0$. Under this
assumption, DM decay is forbidden, but freeze-out through $\chi^0$-$\chi^+$
coannihilation in the early Universe is possible.  Regarding the DM phenomenology
today, neither direct detection nor indirect searches are expected to
yield signals in this coannihilation scenario. At the LHC, however, DM
could be detected in anomalous decays of the ${W_R}$ if $m_{\chi^0} + m_{\chi^+}
< m_{{W_R}}$.  The process of interest is $p p \to {W_R} \to \chi^0 + (\chi^\pm
\to \chi^0 q q')$, where the 3-body $\chi^\pm$ decay proceeds
through an off-shell ${W_R}$.  Possible search channels for this
process are thus jets and missing transverse energy (MET), or the associated production
of a top and bottom quark with MET. Both signatures are plagued by large SM backgrounds,
mainly from vector boson + jets production,
and it is hard to estimate their discovery potential without running full simulations.
In addition, $\chi^+ \chi^-$ pair production would lead to final states involving multiple jets and MET, similar
to typical signatures of supersymmetric models.
Note that in this scenario the branching ratios of the $W_R$ into SM particles
would be reduced compared to our assumptions in the previous sections,
potentially allowing slightly larger values of $\kappa$ to be consistent
with data.

\textbf{c) $Z$- and $Z_R$-mediated DM interactions.}
Since $W'$ bosons typically come with neutral $Z'$-like partners, it is also interesting to consider
$Z'$-mediated DM-SM interactions.  We will do this in the context of the
LRM discussed in Sec.~\ref{sec:LRM}, but our
conclusions easily generalize to other models, in particular to scenarios in
which the LHC diboson anomaly is interpreted as being directly due to a $Z'$
resonance.  The DM candidate could again be one of the standard right-handed
neutrinos $N_R$.  In this case, however, we would again face the problem of
fast DM decay through the ${W_R}$.  Therefore, let us consider a scenario where a
new fermion multiplet $\chi$ with quantum number $(1, 2, -1)$ under $SU(2)_L
\times SU(2)_R \times U(1)_{B-L}$ is added to the model.
The upper (neutral) component of this doublet is the DM candidate $\chi^0$, the
lower (charged) component $\chi^-$ is assumed to be sufficiently heavy for
coannihilations to be negligible.  Since the multiplet $(\chi^0, \chi^-)$ has
the same quantum numbers as the right-handed leptons, there is the possibility
of an undesirable mixing between $\chi^0$ and $N_R$, which would reintroduce
${W_R}$-mediated DM decay.  Such mixing can be forbidden by introducing an
additional symmetry, for instance a dark sector $Z_2$ symmetry.  Note that such
a symmetry does not forbid a Majorana mass term for the $\chi^0$, unlike for
instance a $U(1)$ symmetry.  This is crucial, because if $\chi^0$ was a Dirac
fermion, its vector couplings to the $Z_R$ and (through gauge boson mixing) the
$Z$ would bring it into blatant conflict with direct detection results. For a
Majorana DM particle, however, the dominant effect in direct detection
experiments is spin-dependent scattering through axial vector interactions, for
which limits are much weaker.

Annihilation of the $\chi^0$ in the early Universe proceeds through $s$-channel $Z$
and $Z_R$ exchange into fermionic final states and into $W^+W^-$, with the
fermionic final state dominating.  This implies
in particular that for $m_\chi \sim m_Z / 2$ or $m_\chi \sim m_{Z_R} / 2$ the
total annihilation cross section $\ev{\sigma v_\text{rel}}_\text{tot}$ is
resonantly enhanced. Close to the $Z_R$ resonance, the annihilation cross
section to massless fermions is given by
\begin{align}
  \ev{\sigma v_\text{rel}}(\chi\chi \to f\bar{f}) \simeq
  \frac{n_c v_\text{rel}^2}{6\pi}
    \frac{g_{\chi A}'^2 (g_{f V}'^2 + g_{f A}'^2) m_\chi^2}
         {(4 m_\chi^2 - m_{Z_R}^2)^2 + m_{Z_R}^2 \Gamma_{Z_R}^2} \,.
  \label{eq:sigma-v-ff}
\end{align}
Here $g_{f V}'$, $g_{f A}'$ and $g_{\chi A}'$ are the vector and axial vector
couplings of the final state fermions and the axial vector coupling of the
DM particle to the $Z_R$, respectively.  Explicit expressions
for them have been given in Sec.~\ref{sec:LRM}~\cite{Rizzo:1981dm,Rizzo:1981su}.
Note that $\chi^0$ does not have vector couplings because it is a Majorana
fermion. This the reason for the $v_\text{rel}^2$ suppression in
Eq.~\eqref{eq:sigma-v-ff}.
It can be understood by noting that
due to the Pauli exclusion principle, the two incoming DM particles can only
be in an $s$-wave state if their spins are opposite. The final state fermions
are, however, produced in a spin-1 state due to the chirality structure of
the gauge boson couplings. Thus, either one of them has to experience a
helicity flip (which is only possible for $m_f \neq 0$), or the initial
state DM particles have to be in a $p$-wave state.  Note that in the
resonance region, the $p$-wave contribution proportional to $v_\text{rel}^2$
is dominant. The reason is that, on resonance, an on-shell, spin-1 $Z_R$ boson is produced,
and this requires the DM particles to be in a spin-1 state as well.
Outside the resonance region, the $p$-wave terms dominate in the early Universe,
where $\ev{v_\text{rel}^2} \sim 0.24$, while today, where $\ev{v_\text{rel}^2} \sim 
10^{-6}$ in the Milky Way, it is the helicity-suppressed terms that give the
main contribution.

This implies that the annihilation cross section today is several orders of
magnitude below the thermal relic value, making indirect DM detection in this
scenario extremely challenging.

To compute $\ev{\sigma v_\text{rel}}_\text{tot}$, we have used
FeynCalc~\cite{Mertig:1990an} to evaluate the annihilation cross sections for
the $f\bar{f}$ and $W^+ W^-$ final states.  Note that, in doing so, we need not
only the coupling constants appearing in the simplified expression
Eq.~\eqref{eq:sigma-v-ff}, but also the DM coupling to the SM-like $Z$ boson.
It is given by its coupling to the $Z_R$, multiplied by the $Z$-$Z_R$ mixing
angle $\beta_z (m_Z / m_{Z_R})^2$ as discussed in Sec.~\ref{sec:LRM}.  The
$ZW^+W^-$ coupling is at its SM value, while the $Z_R W^+ W^-$ coupling is
again suppressed by a factor $\beta_z (m_Z / m_{Z_R})^2$.  Note that in
evaluating the cross section for $\chi\chi \to W^+ W^-$, we include only the
transverse polarization states of the internal and external gauge bosons. Thus
we avoid having to include model-dependent diagrams with Higgs boson exchange.
Since annihilation to $W^+ W^-$ is subdominant by a large margin compared to
annihilation to fermions, this approximation will not affect our results.

\begin{figure}
  \centering
    \includegraphics[width=0.6\textwidth]{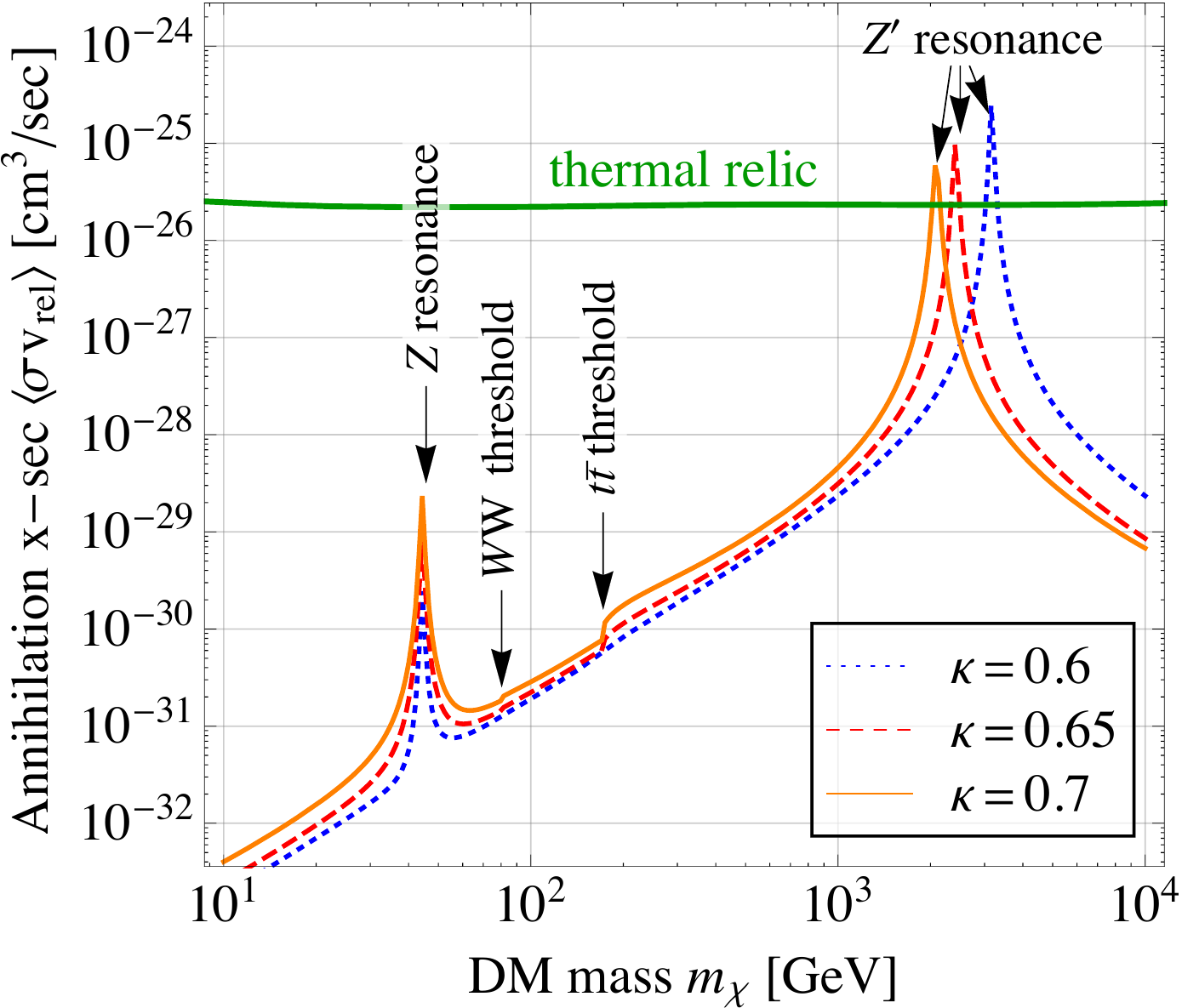}
  \caption{The total DM annihilation cross section
    $\ev{\sigma v_\text{rel}}_\text{tot}$
    as a function of the DM mass for different values of $\kappa = g_R / g_L$.
    We include fermionic final states as well as annihilation to $W^+ W^-$.
    Due to the velocity and helicity-suppression of the annihilation
    cross section, we find that only for $m_\chi \sim m_{Z_R}/2$, the cross section
    can be large enough to avoid DM overproduction in the simplest thermal
    freeze-out scenarios.}
  \label{fig:DM_relic_density}
\end{figure}

We plot $\ev{\sigma v_\text{rel}}_\text{tot}$ as a function of the DM mass in
Fig.~\ref{fig:DM_relic_density} and compare it with the value required for a
thermal relic~\cite{Steigman:2012nb}.  Here we assume the conditions at DM
freeze-out; in particular, we take $\ev{v_\text{rel}^2} \sim 0.24$ for the
average relative velocity of the two annihilating DM particles.  We find that
even at the $Z$ resonance, the helicity and velocity suppression leads to
annihilation cross section several orders of magnitude below the required value
for the correct relic density. Only if the DM mass is close to $m_{Z_R}/2$
mass, the resonant enhancement is large enough to make the annihilation cross
section compatible with the observed relic density.  On the other hand, this
implies that, if the $Z_R$ boson in this scenario is indeed responsible for the
coupling of DM to the SM, the model provides a strong indication for the value
of $m_\chi$. Let us remark again that also an annihilation cross section
somewhat below the naive thermal relic value may be acceptable if the Universe
goes through a phase of extra entropy production after DM freeze-out, thus
diluting the DM density.

\begin{figure}
  \centering
  \includegraphics[width=0.6\textwidth]{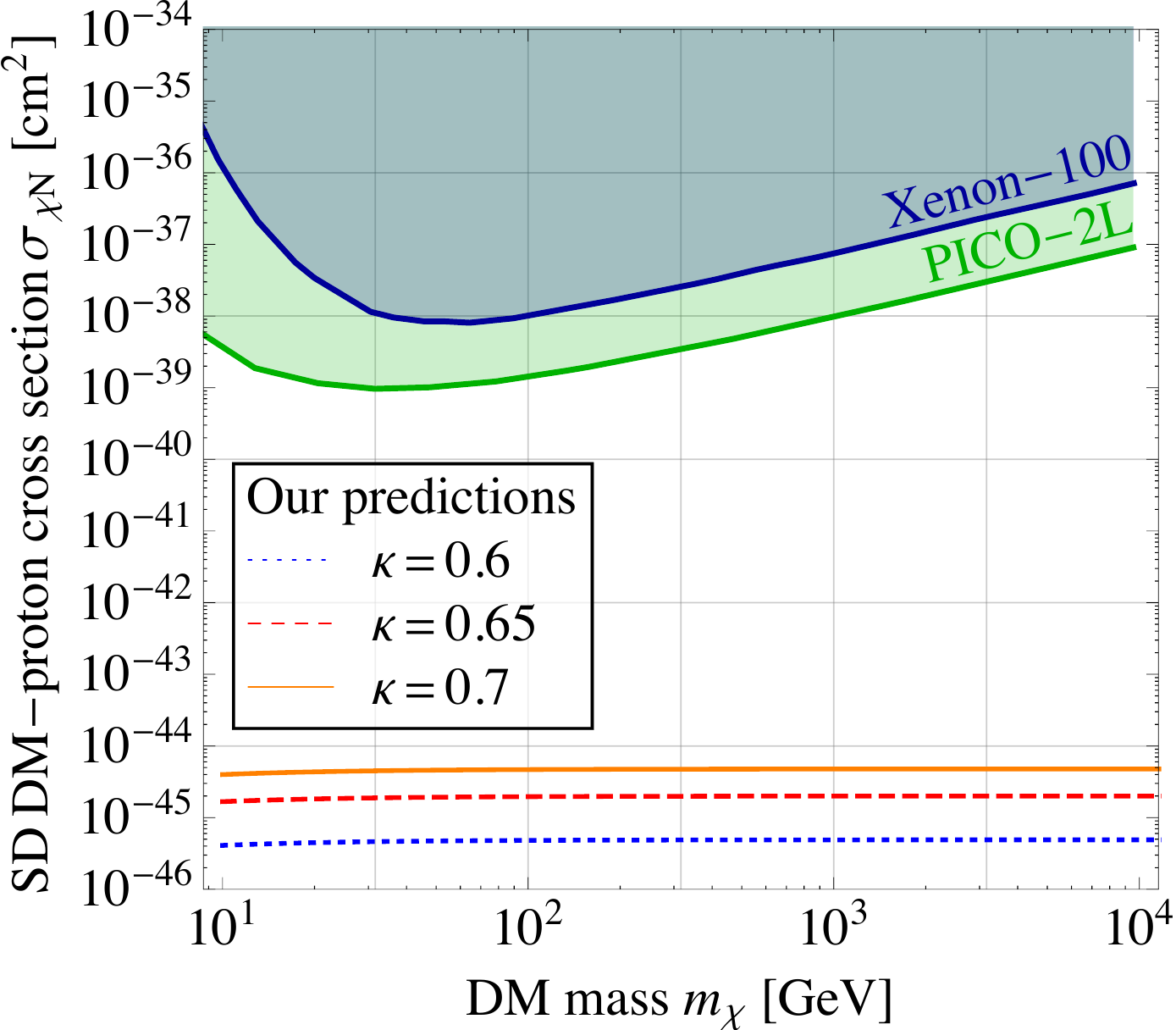}
  \caption{The spin-dependent DM--proton scattering cross section as a function
    of the DM mass for different values
    of $\kappa = g_R / g_L$, compared to spin-dependent direct detection limits
    from the XENON-100~\cite{Aprile:2012nq} and
    PICO-2L~\cite{Amole:2015lsj} experiments.}
  \label{fig:DM_direct_detection}
\end{figure}

It is also important to consider the dark matter--nucleon scattering cross section
probed by direct detection experiments. Since $\chi^0$ is a Majorana particle with
only axial vector couplings, the scattering will be spin-dependent.  The
cross section is
\begin{align}
  \sigma_{\chi N} = \frac{3 m_N^2 m_\chi^2}{\pi (m_N + m_\chi)^2}
  \bigg[ \sum_{q=u,d,s} \Delta_{Nq} \bigg( \frac{g_{qA} g_{\chi A}}{M_Z^2} + 
               \frac{g_{qA}' g_{\chi A}'}{M_Z'^2} \bigg) \bigg]^2 \,,
\end{align}
where $N = p$ or $n$ for scattering on protons and neutrons, respectively.  As
before, primed coupling constants denote couplings to the $Z_R$ and unprimed
ones indicate couplings to the $Z$.  The hadronic form factors $\Delta_{Nq}$
are taken from Ref.~\cite{Belanger:2008sj}.  In
Fig.~\ref{fig:DM_direct_detection} we show the cross section of DM interactions
with protons as a function of the DM mass and of $\kappa$ and compare it to
limits from the XENON-100~\cite{Aprile:2012nq} and PICO-2L~\cite{Amole:2015lsj}
experiments.  Independent of the DM mass, we find that our scenario predicts a
cross section a few orders of magnitude below current exclusion bounds.

We can conclude that this simple scenario gives the correct relic density for
DM close to the $Z_R$ mass, while being fully consistent with current limits
from direct detection experiments.  Unfortunately, it is also not in reach of
these experiments in the foreseeable future. We also expect the $\chi^0$ to be too heavy
to be within the discovery reach of the LHC.

\begin{figure}[htbp]
\centerline{\includegraphics[width=4.5in]{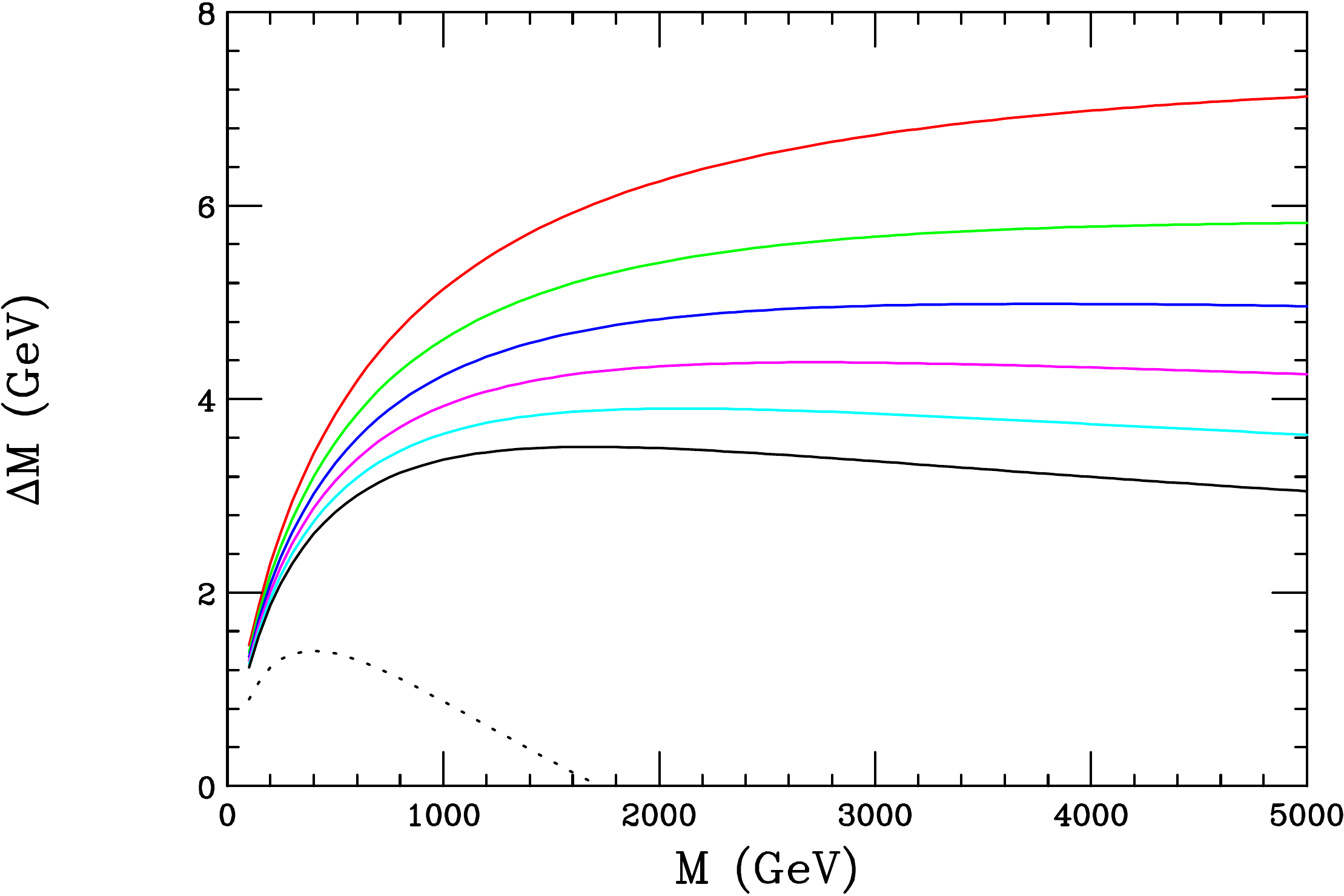}}
\caption{For the minimal left-right DM scenario, we show the radiative mass
  splitting $\Delta M$ between the charged and neutral components of the new
  multiplets $\chi_L$ and $\chi_R$ as a function of the tree level Majorana mass
  $M$. The different colored curves correspond, from top to bottom, to $\kappa$
  values going from 0.56 to 0.71 in steps of 0.03.  We always assume
  $M_{W_R}=1.9$ TeV.  The result for $\kappa=1$ is shown as the dotted curve for
  comparison.}
\label{dm-split}
\end{figure}

\textbf{d) Minimal Left-Right Dark Matter.}
Another possibility for DM in the LRM is to introduce a pair of chiral fermion multiplets 
\begin{equation}
\chi_L \sim (3,1,0), ~~\chi_R \sim (1,3,0)\,,
\end{equation}
which share a common Majorana mass $M$ due to left-right exchange symmetry and whose neutral member(s) can be identified as DM. This scenario was recently considered in 
Ref.~\cite{Heeck:2015qra}. As was discussed there, such a scenario actually leads to a two-component picture for DM. 
Prior to electroweak radiative corrections, all members of the $\chi$ multiplets are degenerate. An issue that arises for the right-handed  component is the \emph{sign} of these corrections, i.\,e., whether or not they drive the masses of the charged 
states below that of that of the neutral member for some choice of the parameter ranges, in particular, the value of $M$. 
Ref.~\cite{Heeck:2015qra} showed that for $M_{W_R}$ = 2 TeV and assuming  $\kappa=1$ one must have $M$ below $\sim 1.8$ TeV, otherwise this mass splitting goes negative thus preventing us from identifying the neutral component as (part of) DM. It 
thus behooves us to determine if this result is robust when we lower the value of $\kappa$ to our range of interest. Employing the $W_R$ and $Z_R$ couplings given above (as well as the general $Z_R-W_R$ mass relationship), we find that the relevant mass splitting is now given by the expression   
\begin{equation}
  {\frac{\Delta M}{M}} = Q^2 ~{\frac{G_F M_W^2}{2\sqrt {2}\pi^2}}~\bigg [ \kappa^2 f(r_{W_R}) - {\frac{\kappa^2-(1+\kappa^2)}{1-x_w}} 
       f(r_{Z_R})-{\frac{x_w^2}{1-x_w}} f(r_Z) -x_w f(r_\gamma) \bigg ]\,,     
\end{equation}
where $r_i=M_i/M$ and $f(r)$ is given by the integral 
\begin{equation}
  \int_0^1 \mathrm{d}x \ 2(1+x) \, \log[x^2+(1-x)r^2]\,,
\end{equation}
which we can evaluate analytically\footnote{It is interesting to observe that the mass splitting vanishes in the case 
when all the mass functions, $f$, are equal as it does in the case of the SM.}. Fig.~\ref{dm-split} shows the values we obtain for $\Delta M$ vs. $M$ as we vary $\kappa$ over the relevant range as well as for the case of $\kappa=1$ for comparison purposes always taking $M_{W_R}$=1.9 TeV. Here we see that the mass splitting is \emph{always} positive for the range of $\kappa$ 
values of interest. We also observe that the magnitude of $\Delta M$ decreases as $\kappa$ value increases. We note that at 
large masses, as $\kappa$ increases, the curves begin to bend downward with an ever increasing slope. For the mass range 
shown here $\Delta M$ always remains positive until values of $\kappa \simeq 0.88$ are reached.   

\textbf{e) DM in supersymmetric Left-Right models.} A fifth possibility of
introducing dark matter in the framework discussed here is to consider
supersymmetric grand unified models based on Left-Right symmetry. 
In this case, the lightest neutralino is an excellent DM candidate, as has been extensively
studied~\cite{Bhattacharya:2013nya,Esteves:2011gk}. However, its phenomenology is
very similar to that of the lightest neutralino in the minimal supersymmetric
standard model (MSSM), with no direct connection to the excesses seen at ATLAS and CMS.

\section{Conclusions}
\label{sec:conclusions}

In this study we have analyzed the observed resonant excesses that appear in different ATLAS and CMS search channels in the invariant mass region
of 1.8\,--\,2.0~TeV. The most prominent of these displays a $3.4\,\sigma$ excess in the search for the hadronic decay of a $WZ$ final state,
though this peak is also sensitive to other diboson states. Both ATLAS and CMS find excesses in the dijet distributions at approximately
the same mass, with a significance around $2.2\,\sigma$ and $1\,\sigma$. There are further potential signals in CMS searches for
semileptonic decays of vector boson pairs as well as for resonances decaying to a $HW$ pair, significant at $1.5\,\sigma$ and $2.2\,\sigma$.
We have investigated potential scenarios that can explain these excesses while being consistent with all
constraints from other searches.

In the first part of our analysis, we have performed a model-independent fit of the cross sections corresponding to the observed and expected event numbers
in all searches from ATLAS and CMS that are sensitive to these and closely related
final states, including the analyses that agree with background expectations. Our fit finds an overall tension between the SM and the data equivalent
to $2.9\,\sigma$. To explain the excess observed in vector boson pair production and associated Higgs production,
a charged resonance is favoured over a neutral resonance.  In particular,  states decaying into $W^+ W^-$ pairs are strongly
constrained from semileptonic searches. The best agreement is found for a charged resonance with approximately equal branching
ratios into $WZ$ and $WH$ pairs, with fitted signal cross sections around 5~fb in each final state. The excess in the dijet distributions
suggest a branching ratio into quarks or gluons that is larger by a factor of at least 10, which is welcome news,
as a sizable coupling of the heavy resonance to quarks or gluons is necessary for a sufficiently large production cross section
at the LHC. Resonance searches in the $tb$ state do not observe an excess, but the upper limit still allows a decay into this
channel of approximately half the size of the dijet branching fraction.

As a next step, we have interpreted these signatures in the context of the Left-Right Symmetric Model based on the extended gauge group
$SU(2)_L\times SU(2)_R\times U(1)'$ as the resonance production of a new 
heavy charged gauge boson, $W_R$. Fitting
this model to the data, we have found that a $W_R$ of 1900~GeV is in good agreement with all analyses, if the right-handed coupling
is in the range $0.35 \lesssim g_R \lesssim 0.45$ and the mixing between the $W$ and $W_R$ is of the order
$\sin \phi_w \sim 0.0015$. This preferred region can be translated into mass constraints for the associated heavy neutral gauge  boson $Z_R$,
for which we find a lower bound of approximately $m_{Z_R} \gtrsim 4$~TeV.
For a lighter $W_R$, the bounds from the fit become stronger, requiring a smaller coupling $g_R$ and a heavier $Z_R$.

In the upcoming 13~TeV run, the LHC will be able to probe this potential signal with very little data. Already with an integrated luminosity of 5~fb$^{-1}$,
the experiments should be able to exclude the dijet signal of our best-fit scenario, followed by sensitivity in the $tb$ channel shortly thereafter
and then in the diboson channels with statistics of roughly 15~fb$^{-1}$.
For a $5\,\sigma$ discovery, we estimate that a luminosity of 20~fb$^{-1}$ is needed in the $jj$ channel, while the $tb$ and especially the
$WZ$ and $WH$ states require more statistics. Whether the 13 TeV LHC will be able to produce the $Z_R$ crucially depends on the 
value of $\kappa$
(or equivalently $m_{Z_R}$). The LHC will begin to probe the interesting parameter space for the new neutral gauge boson with integrated luminosities around 20~fb$^{-1}$,
but it is possible that this gauge boson is too heavy to be accessible at the 13 TeV LHC.

In addition, we have analyzed if this model can simultaneously explain dark 
matter. If the DM annihilates primarily through the $W_R$, we require two new states, one charged and one neutral
(the DM candidate). These would have to be relatively close in mass in order to be able to coannihilate and produce the 
correct relic density. In the case of a $Z_R$ mediator, we introduce a new fermionic doublet with a neutral component 
to act as the dark matter. Here we find that a resonant mechanism is required to produce the correct amount of dark matter 
and thus if this solution is realized in nature, we predict the mass $\sim\frac{1}{2}m_{Z_R}$.

The fact that a number of excesses in different search channels across two experiments can be explained by an existing, simple (and, some might argue, natural)
extension of the Standard Model is exciting. After the first months of data taking at the LHC at 13~TeV we will know more.  We eagerly await
the discovery of symmetry restoration in the upcoming operations of the 13 TeV LHC!

\section*{Note}

While this study was in the last stages of preparation, several other studies appeared 
that analyzed similar models \cite{Gao:2015irw,Cheung:2015nha,Dobrescu:2015qna,Aguilar-Saavedra:2015rna,Hisano:2015gna}. In 
addition, other models have been put forward to explain the 
various excesses \cite{Fukano:2015hga,Franzosi:2015zra,Xue:2015wha,Thamm:2015csa}.

\section*{Acknowledgements}

We would like to express our gratitude to Tilman Plehn for encouragement to study the 
experimental excesses in more detail, as well as for collaboration in the early stages of the project.
We wish to especially thank Gustaaf Brooijmans, David Morse and Chris Pollard for detailed discussions 
regarding the experimental results. We would also like to thank Felix Yu for useful discussions.
In addition, we are grateful to Benjamin Fuks for providing a FeynRules 
implementation of the $W'$ model. The authors would also like to thank the organizers of the Les Houches 
workshop, where part of this work was completed.

J.\,B., J.\,K., and J.\,T.\ would like to thank the kind support of the German Research Foundation (DFG) 
as part of the Forschergruppe `New Physics at the Large Hadron Collider' (FOR 2239), J.\,B.\ also as part of 
the Graduiertenkolleg `Particle physics beyond the Standard Model' (GRK 1940). The work of J.\,H\ and T.\,R.\ was
supported by the Department of Energy, Contract DE-AC02-76SF00515. J.\,K.\ is supported by the DFG
under Grant No.\ \mbox{KO~4820/1--1}.

\appendix
\section{Fit input data}
\label{sec:appendix}

In Sec.~\ref{sec:measurements} we have described a cross-section fit to all ATLAS and CMS analyses sensitive to diboson, $VH$, dijet and $tb$ final states.
This fit is based on a cut-and-count analysis in the mass bins around 1800~GeV. 
In Tbl.~\ref{tab:fit-input} we give the input information our fit uses in more detail, including the exact selections and mass bins that were used
and the number of observed and expected background events in these analyses. Where available, these numbers are based on the papers,
conference notes, supplementary material, and HepData entries published by ATLAS and CMS. Missing pieces of information were roughly estimated
based on the available information. In some cases the estimated signal efficiencies and uncertainties were rescaled to give better agreement with the limits published
by ATLAS and CMS, but we checked that this modification does not affect the overall fit results significiantly.

In analyses involving hadronic decays of gauge bosons, in particular in the ATLAS diboson search~\cite{Aad:2015owa},
$W$ bosons can be reconstructed in the $Z$ selection and vice versa. Based on Fig.~1\,c) in~\cite{Aad:2015owa} we estimate this spill factor to be
\begin{equation}
    \frac {p(\text{$W$ reconstructed as $Z$})} {p(\text{$W$ reconstructed as $W$})}  \approx
    \frac {p(\text{$Z$ reconstructed as $W$})} {p(\text{$Z$ reconstructed as $Z$})}   \approx 0.74 \,.
\end{equation}

The $WW$, $WZ$ and $ZZ$ selections in \cite{Aad:2015owa} are not orthogonal. In fact, in the enlarged signal region there are only 2 observed events that are tagged in
either of the $WW$ or $ZZ$ selections, but not in the $WZ$ category (see Fig.~13 of the auxiliary material published with~\cite{Aad:2015owa}). This is why for
the statistical combination of different searches we follow a conservative approach and only include the results from the $WZ$ selection. 

\begin{table}[tbp]
  \renewcommand{\arraystretch}{1.}
  \begin{scriptsize}
    \begin{tabular*}{1.0\textwidth}{@{\extracolsep{\fill} } llc rrrrr} 
      \multicolumn{8}{l}{\textbf{Fit input data}}   \\ \hline
      Analysis & Selection & Mass bins [GeV] & Obs. & Bkg. & (unc.) & Eff. & (unc.) \\ \hline\hline
      ATLAS $VV$ hadronic \cite{Aad:2015owa} & $WW$ selection & 1750\,--\,2050 & 13 & 8.5 & 1.3 & 0.10 & 0.04 \\ 
      ATLAS $VV$ hadronic  \cite{Aad:2015owa} & $ZZ$ selection & 1750\,--\,2050 & 9 & 3.0 & 0.8 & 0.08 & 0.02\\ 
      ATLAS $VV$ hadronic  \cite{Aad:2015owa} & $WZ$ selection & 1750\,--\,2050 & 18 & 10.0 & 1.5 & 0.09 & 0.03 \\ 
      CMS $VV$ hadronic  \cite{Khachatryan:2014hpa} & Double tagged & 1780\,--\,2030 & 108 & 96.4 & 5.0 & 0.22 & 0.04 \\
      ATLAS $VV$, single lepton \cite{Aad:2015ufa} & Merged region & 1700\,--\,2000 & 8 & 9.1 & 5.2 & 0.27 & 0.01 \\
      CMS $VV$, single lepton \cite{Khachatryan:2014gha} & High purity & 1700\,--\,2000 & 12 & 12.3 & 5.3 & 0.26 & 0.03 \\ 
      ATLAS $VV$, double lepton \cite{Aad:2014xka} & Merged region & 1680\,--\,2060 & 1 & 0.5 & 0.1 & 0.24 & 0.03 \\
      CMS $VV$, double lepton \cite{Khachatryan:2014gha} & High purity & 1700\,--\,2000 & 7 & 3.5 & 0.4 & 0.41 & 0.06 \\
      \hline
      CMS $VH \to b\bar{b}+\nu\ell$ \cite{CMS:2015gla}     & & 1700\,--\,2000 & 3 & 0.5 & 0.4 & 0.06 & 0.01 \\ 
      CMS $VH \to \tau^+\tau^-$ + hadronic vector \cite{Khachatryan:2015ywa}  & & 1500\,--\,2000 & 8 & 8.3 & 3.5 & 0.37 & 0.05 \\
      CMS $VH$, hadronic Higgs \cite{Khachatryan:2015bma} & $bb$ selection & 1690\,--\,2030 & 28 & 27.1 & 4.1 & 0.16 & 0.03 \\
      \hline
      ATLAS dijet \cite{Aad:2014aqa} & & 1706\,--\,2030 & 38326 & 37998 & 90.0 & 0.16 & 0.02 \\
      CMS dijet \cite{Khachatryan:2015sja} & & 1678\,--\,1945 & 114117 & 113438 & 100.0 & 0.38 & 0.04 \\
      \hline
      ATLAS $tb$, hadronic $t$ \cite{Aad:2014xra} & Double tagged & 1600\,--\,2000 & 432 & 410.6 & 28.0 & 0.05 & 0.02 \\
      ATLAS $tb$, leptonic $t$ \cite{Aad:2014xea} & & 1600\,--\,2000 & 14 & 31.5 & 16.9 & 0.06 & 0.02 \\
      CMS $tb$, leptonic $t$ \cite{Chatrchyan:2014koa} & & 1500\,--\,2000 & 178 & 187 & 20.0 & 0.13 & 0.01 \\
      \hline 
    \end{tabular*}
  \end{scriptsize}
\caption{Input data into our fit. We give the observed and expected background events, the uncertainty on the background events,
  the product of all relevant acceptance and efficiency factors, as well as the uncertainty on this number.}
\label{tab:fit-input}
\end{table}


%


\end{document}